\documentstyle[aps,pre,multicol,epsf,subfigure]{revtex}
\def\D{\partial}
\def\grad{\nabla}
\def\div{\nabla \cdot}

\def\Tr{\mbox{Tr}\;}
\def\const{\mbox{const.}}
\def\Av#1{\overline{#1}}
\def\Eq#1{Eq.(\ref{#1})}
\def\Ref#1{(\ref{#1})}
\def\Fig#1{Fig.\ref{#1}}   
\def\eq{\begin{eqnarray}}
\def\qe{\end{eqnarray}}
\def\eqnn{\begin{eqnarray*}}
\def\qenn{\end{eqnarray*}}
\def\nn{\nonumber}
\def\be{\bm{e}}

\def\bn{\bm{n}}
\def\bq{\bm{q}}
\def\br{\bm{r}}
\def\bu{\bm{u}}

\def\simge{\;\lower3pt\hbox{$\stackrel{\textstyle >}{\sim}$}\;}
\def\simle{\;\lower3pt\hbox{$\stackrel{\textstyle <}{\sim}$}\;}
\def\bm#1{\mbox{\boldmath $#1$}}
\def\tensor#1{{\sf #1}}
\def\lrL#1{\left[#1\right]}

\def\lrS#1{\left(#1\right)}
\def\lrF#1{\left|#1\right|}
\def\lrA#1{\left\langle #1 \right\rangle}

\def\biglrL#1{\bigl[ #1 \bigr]}

\def\biglrS#1{\bigl( #1 \bigr)}
\def\biglrF#1{\bigl| #1 \bigr|}
\def\biglrA#1{\bigl\langle #1 \bigr\rangle}
\def\bigglrL#1{\biggl[ #1 \biggr]}

\def\bigglrS#1{\biggl( #1 \biggr)}
\def\bigglrF#1{\biggl|#1\biggr|}
\def\bigglrA#1{\biggl\langle #1 \biggr\rangle}
\def\mycomment#1{}
\def\mycaption#1{\nopagebreak[4]\hspace{0mm}\\\begin{minipage}[h]{85mm}\caption{#1}\end{minipage}\vfill}
\def\myfigure#1#2{\nopagebreak[4]\hspace{0mm}\\\begin{minipage}[h]{85mm}#1\caption{#2}\end{minipage}\vfill}
\def\f#1#2{\frac{#1}{#2}}
\def\der#1#2{\f{\D #1}{\D #2}}
\def\fder#1#2{\f{\delta #1}{\delta #2}}

\newcommand{\tQ}{\tensor{Q}}

\newcommand{\tR}{\tensor{R}}
\newcommand{\tP}{\tensor{P}}
\newcommand{\tG}{\tensor{G}}
\newcommand{\tL}{\tensor{L}}

\newcommand{\tC}{\tensor{C}}
\newcommand{\tI}{\tensor{I}}
\newcommand{\tLambda}{\tensor{\Lambda}}

\newcommand{\qh}{\hat{q}}
\newcommand{\bqh}{\hat{\bq}}

\newcommand{\bg}{\bm{g}}
\newcommand{\brho}{\bm{\rho}}
\newcommand{\bR}{\bm{R}}
\newcommand{\gnuplotdiamond}{\diamond\hspace{-1.34mm}\cdot\,}

\title{
Soft and non-soft structural transitions 
in disordered nematic networks
}
\author {Nariya Uchida~\thanks{Electronic address : uchida@ton.scphys.kyoto-u.ac.jp}}
\address{Department of Physics, Kyoto University, Kyoto 606, Japan}
\date   {March 31, 2000}
\begin{document}
\draft

\bibliographystyle{prsty}

\maketitle
\begin{abstract}
Properties of disordered
nematic elastomers and gels are theoretically
investigated with emphasis on
the roles of non-local elastic interactions
and crosslinking conditions.
Networks originally crosslinked
in the isotropic phase lose their
long-range orientational order
by the action of quenched random stresses, 
which we incorporate into the affine-deformation
model of nematic rubber elasticity.
We present a detailed picture of
mechanical quasi-Goldstone modes,
which accounts for 
an almost completely soft polydomain-monodomain (P-M) transition
under strain
as well as a ``four-leaf clover'' pattern
in depolarized light scattering intensity.
Dynamical relaxation of the domain structure
is studied using a simple model.
The peak wavenumber of the structure factor
obeys a power-law-type slow
kinetics and goes to zero 
in true mechanical equilibrium.
The effect of quenched disorder 
on director fluctuation
in the monodomain state is analyzed. 
The random frozen contribution 
to the fluctuation amplitude
dominates the thermal one,
at long wavelengths
and near the P-M transition threshold.
We also study networks obtained by crosslinking
polydomain nematic polymer melts. 
The memory of initial director configuration
acts as correlated and strong quenched disorder,
which renders the P-M transition non-soft.
The spatial distribution 
of the elastic free energy 
is strongly dehomogenized by external strain,
in contrast to the case of isotropically 
crosslinked networks.

\end{abstract}


\begin{multicols}{2} 

\section{Introduction}

Elastomers and gels are intrinsically disordered solids
that retain the memory of their initial states.
The non-equilibrium nature of their
fabrication processes causes frozen heterogeneities
in the network structure, which range in size from
mesoscopic to macroscopic scales~\cite{Dusek-Prins,Shibayama}.
The presence of the quenched disorder comes to the fore
when we introduce some soft order in the system.
For instance, density fluctuations
of swollen gels near the critical point
are strongly enhanced by the 
heterogeneities.
Under stretching, they produce
the so-called ``abnormal butterfly'' 
pattern in small-angle neutron scattering 
intensity~\cite{Mendes-etal,Bastide-etal,Onuki_butterfly,Bastide-Candau}.
It illustrates how
the elasticity of gels gives rise to
a non-trivial effect unexpected
in other random systems.
Here we address another example of soft order
in disordered elastic networks.

Nematic liquid-crystalline elastomers and gels constitute 
a unique class of solids 
characterized by a coupling between
the orientational and translational degrees of freedom.
Physical consequences of the strain-orientation coupling
have been the subject of a considerable 
amount of studies, both theoretical and experimental.
Notable theoretical advances in the past include:
(i) De Gennes~\cite{deGennesCRAS}
showed that a spontaneous elongation along the director
is induced by the isotropic-nematic (I-N) transition;
(ii)
A molecular model of nematic networks
was constructed by Warner et al.~\cite{WarnerOriginal},
extending the classical affine-deformation model
of rubber elasticity;
(iii)
Uniformly oriented networks possess soft modes of 
strain and orientation fluctuations
that do not accompany any change of 
rubber-elastic free energy.
It was first predicted by Golubovi\'c and 
Lubensky~\cite{Golubovic-Lubensky}  
on a phenomenological ground
and later extended 
by the affine-deformation theory~\cite{WarnerSoft,Olmsted}.
Thus and in other ways, the behavior of homogeneous and clean
nematic networks is now fairly well understood~\cite{Warner-TerentjevReview}.

In practice, however, nematic networks in equilibrium 
quite often 
exhibit polydomain director textures, 
where the orientational correlation 
length is typically in the micron range.
Under external strain, polydomain 
networks undergo a 
structural change into a macroscopically 
aligned monodomain state,
where the director lies along the 
extensional direction.
This change, called the polydomain-monodomain (P-M) 
transition, is characterized by a highly non-linear
mechanical
response~\cite{Finkelmann89,Finkelmann94,Bergmann-etal,Clarke2,Zubarev-etal,Clarke-Terentjev,Brand-FinkelmannReview}.
The strain-stress curve shows a small slope in 
the partially aligned (polydomain) state.
Depending on the material and the 
method of synthesis, the slope is sometimes 
vanishingly small while it is sizable
in other cases.
The macroscopic stress as a function of strain
shows a steep rise
as the system turns into the monodomain state.

There have been 
a few theoretical attempts 
to describe polydomain networks 
and their mechanical responses.
Ten Bosch and Varichon~\cite{tenBosch-Varichon} 
set up the first model,
in which they attributed 
the origin of the equilibrium texture
to a random anchoring field exerted
by network crosslinks. 
An interesting 
analogy with random anisotropy magnets~\cite{Clarke1} was
pursued by
Fridrikh and Terentjev~\cite{Fridrikh-Terentjev}.
They proposed a mapping to the XY model under random 
and homogeneous magnetic fields,
from analysis of which they 
predicted a discontinuous 
stress-orientation curve.

Nonetheless, the role of
strain-orientation coupling 
in polydomain networks is still far from clear.
There are two aspects to be considered.
Firstly, the previous theories
assume only local interactions between domains,
for instance by arguing that the elastic 
free energy localizes in domain walls
under strain~\cite{Fridrikh-Terentjev}.
In general, however,
inhomogeneities in an elastic material
cause non-local or long-range
interactions mediated by the strain field.
Such elastic interactions control the physics
of various systems, such as
solids with dislocations~\cite{LandauElasticity}
or surface defects~\cite{Lau-Kohn,Marchenko-Parshin},
phase separating alloys~\cite{CahnAlloy,OnukiAlloy},
gels undergoing swelling~\cite{TanakaT,OnukiGel}, and
membranes with inclusions~\cite{Goulian-etal}.
Disordered nematic networks provide another intriguing 
example, and differ from any of the above materials 
in having  a non-scalar order parameter.
Secondly, 
the mechanical response should strongly
depend on the crosslinking condition.
Polydomain elastomers have been obtained by either 
of the following ways~\cite{Finkelmann94,Zubarev-etal};
(i) to crosslink a polymer melt in the isotropic phase
and then cool it into the nematic phase;
(ii) to crosslink a nematic polymer melt containing polydomain textures.
These two cases have not been theoretically well distinguished so far.
We shall refer to them as the cases of isotropic and 
anisotropic crosslinkings, respectively.

Recently,
we studied the elastic interaction in
isotropically crosslinked networks~\cite{Uchida-Onuki},
and found an almost completely soft P-M transition~\cite{Uchida}.
The macroscopic stress due to the strain-orientation coupling
was shown to be slightly negative and of $O(\alpha^2)$
in the P-M transition regime,
where $\alpha$ is the degree of chain anisotropy.
This contrasts with the earlier prediction of
a positive stress of $O(\alpha)$~\cite{Fridrikh-Terentjev}.
The elastic interaction also produces
a ``four-leaf clover'' pattern
in the depolarized light scattering intensity, 
which resembles the experimental 
observation by Clarke et al.~\cite{Clarke2,Clarke1}.

In this article, we
extend previous work~\cite{Uchida-Onuki,Uchida}
and provide the details of our picture of
the P-M transition.
Here let us summarize the ideas and results which 
we have not emphasized in previous work.
Firstly,
we pursue the idea
that random internal stresses
destroy the long-range orientational order,
which was suggested (but not proven) earlier
in a broader context~\cite{Golubovic-Lubensky}.
This will be done by
incorporating the notion of
frozen heterogeneous strains into
the extended affine-deformation theory~\cite{WarnerOriginal}.
We argue that the random internal stresses
act as stronger sources of
disorder than the random molecular field
due to crosslinks~\cite{tenBosch-Varichon,Clarke1}.
Secondly,
evolution of domain structure with and without external 
stretching is numerically simulated
by a simplified dynamical model.
The ``four-leaf clover'' scattering pattern has
four peaks at finite wavenumbers, and
the peak height is a non-monotonic
function of the macroscopic strain,
in qualitative agreement with experiment.
We find a slow dynamical relaxation of
the structure factor, and show that the peak wavenumber
asymptotically goes to zero in the long-time limit.
Thirdly,
we study the case of anisotropic crosslinking.
In this case, the initial director configuration
of a macroscopic polydomain texture is memorized 
into the network. It provides a source of
strong and correlated disorder,
resulting in a non-soft P-M transition.
The spatial distribution of elastic free energy 
in anisotropically 
crosslinked networks 
is strongly dehomogenized by strain,
while that in
isotropically crosslinked networks
is unchanged during the P-M transition.

This paper is organized as follows.
In Section II, we introduce a random stress
model, derive an effective free energy,
and discuss the mechanism of the soft mechanical response.
Section III describes a numerical simulation of
the polydomain state and the P-M transition.
In Section IV, we analyze the effect of random stresses
on director fluctuations in the monodomain state.
We study networks prepared in the nematic phase in Section V.
In Section VI, we summarize the results in comparison 
to existing experiments, and conclude with a proposal of
future directions.

\section{Model and Analysis} 

\subsection{Random stresses in isotropic networks}

\begin{figure}
\epsfxsize=120pt \epsffile{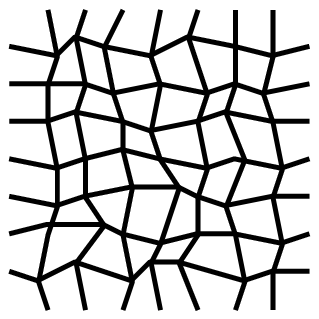}
\mycaption{
Schematic illustration of disordered
network structure.
When the nematic order is introduced,
the director is preferentially oriented 
along the extensional axes of frozen network strains.
\label{mesh}
}
\end{figure}

It is known since long ago~\cite{Dusek-Prins,deGennesPolymerBook} 
that the network structure of gels are 
often heterogeneous on many length-scales,
which are considerably larger than
the mesh size (see \Fig{mesh}). 
In the swollen state, these imperfections
manifest themselves as density inhomogeneity and 
are observed through the so-called butterfly
pattern in neutron scattering intensity
or as speckles in light scattering
experiment~\cite{Shibayama,Bastide-Candau}.
Although less discussed in the literature,
it is natural to expect that elastomers, 
often fabricated by drying gels,
also contain the memory of heterogeneous network 
formation.
The frozen heterogeneities reflect
the non-equilibrium nature of the crosslinking processes,
and produce random internal stresses in the material.
While the roles of random stresses in gels and other
amorphous solids have been discussed from a phenomenological 
point of view~\cite{Golubovic-Lubensky,Alexander},
much remains to be done to understand
them on the basis of a molecular theory~\cite{Panyukov-Rabin}.
In this subsection,
we recapitulate the notion of random stresses 
using the classical affine-deformation theory
of isotropic rubber networks~\cite{FloryBook},
in order to prepare for modeling disordered 
nematic networks in the next subsection.

The basic object in the affine-deformation theory 
is the probability distribution
of the chain's end-to-end vector $\brho$.
The distribution function at thermal equilibrium
is isotropic and Gaussian, and given by 
\eq
P_{eq}(\brho) = {\cal N}^{-1} \exp \bigglrS{-\f{d}{2\Omega} \rho^2},
\label{Peq}
\qe
where $\Omega$ $=\lrA{\rho^2}_{eq}$
is a constant,
$d$ is the spatial dimension,
and ${\cal N}=\int d\brho  P_{eq}(\brho)$
is the normalization factor.
The macroscopic deformation of the network
is described by the Cauchy deformation tensor,
\eq
\Lambda_{ij} = \der{r_i}{r_{0j}}, 
\qe
where
$\br$ and $\br_0$ are the positions of material points
at observation and at the moment of crosslinking, respectively.
The basic assumption of the theory is that
each chain's end-to-end vector affinely changes
as $\brho \to \tensor{\Lambda} \cdot \brho$
in response to the macroscopic deformation.
The free energy per chain is given by~\cite{FloryBook}
\eq
f_{chain} = - k_BT \int d\brho \; P_0(\brho) \ln P_{eq}(\tLambda \cdot \brho),
\label{fchain}
\qe
where $P_0(\brho)$ is
the probability distribution function
at the moment of crosslinking,
which is not necessarily
identical to the equilibrium distribution.
We assume that the chains are distorted
before crosslinking, and
denote the deviation from the equilibrium conformation
by a tensor $\tR$, defined by
\eq
\lrA{\brho \brho}_0 = \Omega (\tI + \tR),
\label{defR}
\qe
where $\tI$ is the unit tensor.
If the deviation is not very large and 
the chains are not stretched out,
we may still approximate $P_0$ to be Gaussian,  and put
\eq
P_0(\brho) = {\cal N}_0^{-1}\exp\bigglrL{
\f{d}{2\Omega}\brho \cdot(\tI+\tR)^{-1} \cdot \brho
}.
\label{P0}
\qe
Substituting (\ref{Peq}) and (\ref{P0})
into \Eq{fchain},
we have 
\eq
f_{chain} = \f{k_BT}{2} \bigglrL{ \Tr \tG + \tR : \tG +
\ln \det (\tI + \tR) 
},
\label{fchain2}
\qe
where $G_{ij} = \Lambda_{ki}\Lambda_{kj}$
is the metric tensor of deformation.
The equation (\ref{fchain2}) is not new and
essentially contained in the classical theory of Flory~\cite{Flory}.
Taking the spatial heterogeneity of $\tR$
into account and
neglecting terms 
independent of $\tensor{\Lambda}$,
the total elastic free energy is written as
\eq
F_{el} = \f{k_BT \nu_0}{2}
\int d\br_0  \lrS{\Tr{\tG} + \tR : \tG},
\label{Feliso}
\qe
where $\nu_0$ is the number density of subchains.
Inhomogeneous contribution of the form $\tR : \tG$
can be also derived from Cauchy's theory  
of solids bound by a central force~\cite{Alexander,Love}.
We shall call $R_{ij}$ the quenched random stress~\cite{Golubovic-Lubensky},
although it is more directly related to
quenched random {\it strain} in the present model.
For simplicity,
we assume that the frozen heterogeneities
have a single characteristic size $\xi_R$,
which is substantially larger than the mesh size.
After a coarse-graining on the scale $\xi_R$, we can regard
$R_{ij}$ as a spatially uncorrelated Gaussian 
random variable satisfying
\eq
\lrA{R_{ij}(\br_0)} &=& 0,
\label{avR}
\\
\lrA{R_{ij}(\br_0) R_{kl}(\br'_0)} &=& 
\xi_R^d \delta(\br_0 - \br'_0)
\cdot
\nn\\&&\hspace{-20mm}
\bigglrL{
\beta^2 \bigglrS{
\delta_{ik} \delta_{jl} + \delta_{il} \delta_{jk}  
- \f2d \delta_{ij} \delta_{kl} 
}
+
\beta'^2 \delta_{ij} \delta_{kl}
}.
\label{corrR}
\qe
The dimensionless constants $\beta$ and $\beta'$
represent the
magnitudes of shear and dilatational quenched  strains,
respectively.

\subsection{Random stresses in nematic networks}

Next we consider nematic elastomers and gels.
Warner et al.~\cite{WarnerOriginal} constructed
an affine-deformation theory of nematic networks,
by generalizing the classical theory.
Their basic observation is that
nematic chains with low backbone rigidity are 
well characterized by an anisotropic Gaussian 
conformation, elongated along the director. 
The equilibrium 
distribution of the end-to-end vector 
can be written in the form,
\eq
P_{eq}(\brho) &=& {\cal N}'^{-1}
\exp \bigglrL{ -\f{d}{2 \Omega'} \; \brho \cdot (\tI - \alpha \tQ)
\cdot \brho},
\label{Pnematic}
\qe
where $\alpha$ is the degree of chain anisotropy and
\eq
Q_{ij} = Q_0 \bigglrS{\delta_{ij} - \f1d n_i n_j}
\qe
is the orientational order parameter with $\bn$ being the director.
We consider a system deep in
the nematic phase and put $Q_0=1$;
the state of orientation is completely 
specified by the director.
The coupling constant $\alpha$
is expressed in terms of 
the parameters used in~\cite{Warner-TerentjevReview} as
\eq
\alpha = 
\f{\ell_\parallel - \ell_\perp}{(1-1/d)\ell_\parallel + (1/d)\ell_\perp}.
\label{alpha}
\qe
Note that $\alpha$ does not exceed $d/(d-1)$,
the value attained
in the anisotropic limit $\ell_{\parallel}/\ell_{\perp}\to\infty$.
An advantage of the affine-deformation model
is that it can describe arbitrary crosslinking conditions;
the networks can be fabricated either
in the isotropic or the nematic phase.
First we consider networks originally
crosslinked in the isotropic phase,
and shall treat the case of anisotropic
crosslinking in Section V.
The random stresses are now readily incorporated 
into the original model.
For the case of isotropic crosslinking,
the initial chain conformation can be described by \Eq{P0}, 
with (\ref{avR}) and (\ref{corrR}). 
Substituting (\ref{P0}) and (\ref{Pnematic}) into \Eq{fchain},
and dropping terms independent of $\tensor{\Lambda}$ and/or $\tQ$, we 
arrive at the elastic free energy,
\eq
F_{el} = \f{\mu}{2} \int d\br \;
\Tr \bigglrL{
(\tI + \tR) \cdot \tLambda^T \cdot (\tI - \alpha \tQ) \cdot \tLambda - \tI
}
\label{Felbasic}
\qe
with $\mu=k_BT\nu_0(\Omega/\Omega')$.
Here we subtracted a constant
so that $F_{el}$ vanishes when
$\tensor{\Lambda}=\tI$ and
$\alpha=\beta=0$.
We also 
replaced $\int d\br_0$ with $\int d\br$,
assuming an incompressible network 
and imposing the local constraint,
\eq
\det \tLambda = 1.
\label{incompressible}
\qe
We decompose the elastic free energy into 
proper and disorder contributions as
$F_{el} = F_{el}^P + F_{el}^D$,
where, by definition, the former is given
by formally putting $\tR=0$ in the right hand side of \Ref{Felbasic},
and
\eq
F_{el}^D = \f{\mu}{2} \int d\br \;
\Tr \bigglrL{
\tR \cdot \tLambda^T \cdot (\tI - \alpha \tQ) \cdot \tLambda
}.
\qe
The total free energy of the system is written as $F= F_{el} + F_F$
where $F_F$ is the Frank free energy, for which
we use the so-called one-constant approximation~\cite{deGennesLCbook}, 
\eq
F_F = \f{K}{2}\int d\br(\grad \bn)^2.
\qe
We assume that the average deformation $\Av{\Lambda}_{ij}$
is a uniaxial strain along
the $x$-axis, parameterized by the elongation ratio $\lambda$,
as
\eq
\Av{\tLambda} &=& \lambda \be_x \be_x 
+  \lambda^{-1/(d-1)} (\tI-\be_x \be_x).
\label{averagestrain}
\qe
The internal displacement is defined  as the deviation from 
the average deformation,
\eq
\bu = \br - \Av{\Lambda} \cdot \br_0.
\label{displacement}
\qe
with which the deformation tensor is expressed as
\eq
\Lambda_{ij} = \Av{\Lambda}_{kj} (\delta_{ki}+ \D_k u_i)
\label{Lambda}
\qe
(here and hereafter, we imply summation over repeated indices $i$,$j$,$k$,$l$, 
and $m$).

\begin{figure}
\epsfxsize=155pt \epsffile{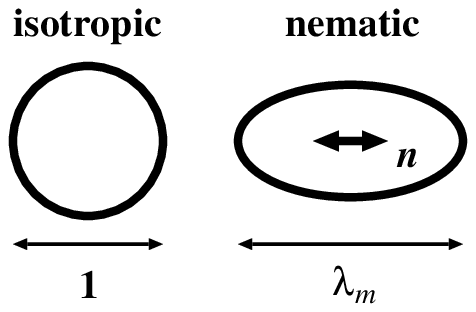}
\mycaption{
Spontaneous macroscopic elongation of an ideal (clean) nematic gel,
induced by the I-N transition.
\label{spont}
}
\end{figure}

In the absence of quenched disorder,
the elastic free energy is minimized at
$\lambda=\lambda_m$ and $\bu=0$, where
\eq
\lambda_m = \bigglrL{\f{1+(1/d)\alpha}{1-(1-1/d)\alpha}}^{(d-1)/2d}
\label{lambdam}
\qe
is the ratio of the spontaneous elongation
induced by the isotropic-nematic
transition~\cite{deGennesCRAS,Warner-TerentjevReview} (see \Fig{spont}).
However, if the random stresses are enough strong,
the long-range orientational order is destroyed and
the ground state of the system is shifted to a polydomain state
with $\lambda=1$,
as we shall see.
Hereafter and throughout the paper,
we regard $\lambda$ as an externally controlled parameter.

\subsection{Effective free energy}

In this subsection, we derive
the effective free energy in the
mechanical equilibrium state
under the constraint $\lambda=1$,
and discuss its physical consequences.
Substituting
Eqs.(\ref{averagestrain}), (\ref{displacement}) and (\ref{Lambda})
into \Eq{Felbasic},
expanding it with respect to $\grad\bu$
and retaining 
a bilinear form in $\grad\bu$, $\tR$ and $\tQ$,
we have
\eq
F_{el} \big|_{\lambda=1} &=& 
\f{\mu}{2} \int d\br
\bigglrL{ (\D_i u_j)^2
+ 2 (R_{ij} - \alpha Q_{ij})\D_i u_j  
\nn\\&&\hspace{30mm}
-  \alpha R_{ij} Q_{ij} 
}
\label{basicDu}
\qe
From this we eliminate the displacement field 
using
the mechanical equilibrium condition,
\eq
\fder{F_{el}}{\bu}=0,
\label{equilibrium}
\qe
and the incompressibility condition \Ref{incompressibility}, which 
to the lowest order in $\grad\bu$ reads
\eq
\div \bu = 0.
\label{incompressibility}
\qe
After a straightforward calculation following
the procedure described in~\cite{Uchida-Onuki},
we obtain an effective elastic free energy
which is correct to the quadratic order in $\alpha$, $\beta$, 
and $\beta'$,
as
\eq
\tilde{F}_{el} \big|_{\lambda=1}
&=& - \f{\mu}{2} \int_{\bq}
\bigglrL{
\bigglrF{
(\tI- \bqh \bqh) \cdot 
\bigglrS{
\bqh \cdot \tR(\bq) - \alpha \bqh \cdot \tQ(\bq) }
}^2 
\nn\\
&& \hspace{20mm} + \alpha \tR(\bq) : \tQ(-\bq)
},
\label{FeldD}
\qe
where $\bqh = \bq/|\bq|$ and $\int_{\bq} = \int (2\pi)^{-d}d\bq$
(the tilde is to put to express the effective nature
of the free energy).
The proper contribution to the
free energy is given by
\eq
\tilde{F}_{el}^P 
&=& - \f{\mu\alpha^2}{2} \int_{\bq}
\bigglrF{
(\tI- \bqh \bqh) \cdot 
\biglrS{
\bqh \cdot \tQ(\bq) 
}
}^2.
\label{Felproper}
\qe
In the real space,
\Eq{Felproper} is rewritten 
in the form of a two-body long-range 
interaction, as
\eq
\tilde{F}_{el}^P &=& - \f{\mu\alpha^2}{2} \int d\br \int d\br' 
\nn\\ &&
\biggl[ 
Q_{ik}(\br) \cdot \D_i \D_j G_1(\br-\br') \cdot Q_{jk}(\br')
\nn\\ && 
+ Q_{ij}(\br) \cdot \D_i \D_j \D_k \D_l G_2(\br-\br') \cdot Q_{kl}(\br') 
\biggr],
\label{Green}
\qe
where $G_n(\br) \; (n=1,2)$ are the Green functions
defined by
\eq
(\grad^2)^n G_n(\br) &=& - \delta(\br),
\\
G_{n} (r \to \infty) &=& 0.
\qe
In a similar manner, the disorder part of 
the free energy is  written as
\eq
\tilde{F}_{el}^D &=&  \mu \alpha \int d\br \int d\br' 
\nn\\ && 
\biggl[ 
R_{ik}(\br) \cdot 
\bigglrS{
-\f12 \delta_{ij} \delta(\br-\br') +
\D_i \D_j G_1(\br-\br') 
}
\cdot Q_{jk}(\br')
\nn\\ && 
+ R_{ij}(\br) \cdot \D_i \D_j \D_k \D_l G_2(\br-\br') \cdot Q_{kl}(\br') 
\biggr] + \const,
\label{GreenR}
\qe
In a given direction of $\bR=\br-\br'$,
the kernels $\D_i\D_j G_1$ and $\D_i \D_j \D_k \D_l  G_2$
in \Ref{Green} and \Ref{GreenR} decay in proportion to $R^{-d}$.

Let us discuss the physical meaning of the effective
free energy.
First we consider the disorder part,
neglecting the proper elastic
interaction for the moment.
If we decompose the random stress
into the dilatational part $R_{kk} \delta_{ij}/d$
and the shear (traceless) part $R_{ij}-R_{kk} \delta_{ij}/d$,
the former
makes no contribution to the free energy \Ref{basicDu}
because of \Ref{incompressibility}
and the tracelessness of $\tQ$.
Thus, only the shear portion of the random stresses
(whose strength is represented by $\beta$)
is relevant, at least in the bilinear order.
It is intuitively obvious that a mere volume change 
does not create any preferential director orientation,
while anisotropic strain does.
As seen from \Eq{GreenR},
the random stresses both locally and non-locally
act on the director field.
The classical scaling argument by Imry and Ma~\cite{Imry-Ma}
tells that
arbitrary weak random stresses destroy
the long-range orientation order
in dimensions lower than four~\footnote{
Although the original Imry-Ma argument assumes
an uncorrelated random field,
it is easy to see that it also holds in the present case,
including the scaling law for the domain size.
To see this, it is useful to
rewrite the right hand side of \Eq{GreenR} 
into the form $\mu \alpha \int d{\bf r} \tP({\bf r}):\tQ({\bf r})$,
where the effective random field $\tP$
has a long-range correlation schematically
represented as
\eq
\lrA{P_{ij}({\bf r}) P_{kl}({\bf r'})} = \xi_R^d \biglrL{
\Pi_{ijkl} \delta({\bf r-r'}) 
+ \Pi'_{ijkl} |{\bf r - r'}|^{-d} }, 
\qe
where $\tensor{\Pi'}$ depends on the direction of ${\bf r-r'}$
but not on its magnitude.
Since both $\tensor{\Pi}$ and $\tensor{\Pi'}$ are dimensionless
quantities, there appears no
additional characteristic length that
affects the Imry-Ma scaling of disorder free energy contained
per domain.
}.
The domain size or the orientational correlation length,
which we denote by $\xi$,
is determined by a balance between
the effects of random stresses and Frank elasticity.
The effective strength of disorder
is expressed by the dimensionless parameter,
\eq
D = \f{\mu \alpha \beta}{K} \cdot \xi_R^2.
\label{disorder_strength}
\qe
According to the Imry-Ma argument,
the domain size scales as $\xi/\xi_R \propto D^{2/(d-4)}$
in the weak disorder regime $D\ll1$.
For a strong disorder $D \simge 1$,
we should have 
$\xi \sim \xi_R$ and 
optimization of the director field
will reduce
the disorder free energy
density roughly by $\mu \alpha \beta$.

Next we turn to the proper
elastic interaction.
It should play only a secondary role in
selecting the domain size $\xi$
because of the invariance of \Eq{Felproper}
against a change of scale $\bq \to \const \times \bq$.
However, it creates a characteristic anisotropy in the
the orientational correlation.
We see this first in the two dimensional case.
In 2D, the orientational configuration is specified by
the director's azimuthal angle $\theta=\theta(\br)$, as
\eq
\bn = (\cos \theta, \sin \theta),
\qe
or, equivalently,
\eq
\tQ = \f12
\left[
\begin{array}{cr} 
\cos 2 \theta & \sin 2 \theta \\
\sin 2 \theta &  - \cos 2 \theta
\end{array} \right].
\qe
A straightforward calculation reduces
\Eq{Green} to 
\eq
\tilde{F}_{el}^P &=& \f{\mu\alpha^2}{16 \pi} \int d\br \int d\br' 
\f{1}{R^2}  
\nn\\&&\quad
\cdot \cos \bigglrL{ 2(\theta(\br)-\psi) + 2 (\theta(\br')-\psi)},
\label{Green2D}
\qe
where $\psi$
is the azimuthal angle of 
$\bR = \br - \br' = 
|\bR|(\cos \psi , \sin \psi)$.
From the angle-dependence of the integrand,
we expect that 
the above free energy is minimized by
a ``checkered'' domain configuration
as depicted in \Fig{preferN2D}.
Correlation in directions
parallel and perpendicular
to the local director is suppressed while
those in oblique directions are enhanced.
It has the following simple interpretation.
Upon the isotropic-nematic transition,
each part of the network tends
to elongate along the local director.
The domain in the center of the figure
pushes the top neighbor upward,
pulls the left neighbor rightward, and so on.
To reduce the mechanical conflict
without violating the global constraint $\lambda=1$,
the top and left domains are reoriented perpendicular
to the central one.
This domain reconfiguration
enables the I-N transition-induced elongation 
along the local director,
despite of spatial inhomogeneity.

The same picture holds for orientational correlation
in three dimensions.
In 3D, \Eq{Green} becomes
\eq
\tilde{F}_{el}^P &=& \f{\mu\alpha^2}{16 \pi} \int d\br \int d\br' 
\f{1}{R^3} \; g(\bn, \bn', \hat{\bR}), 
\\
g(\bn, \bn', \hat{\bR}) &=&
- \f53 
+ 4 (\bn \cdot \bn')^2 
+ (\bn \cdot \hat{\bR})^2 
+ (\bn' \cdot \hat{\bR})^2 
\nn\\ 
&-& 18 (\bn \cdot \bn')(\bn \cdot \hat{\bR})(\bn' \cdot \hat{\bR})
\nn\\ 
&+& 15 (\bn \cdot \hat{\bR})^2 (\bn' \cdot \hat{\bR})^2
,
\qe
where $\bn=\bn(\br)$, $\bn' = \bn(\br')$ and $\hat{\bR} = \bR / |\bR|$.
Correlation in the direction
parallel to the director
is suppressed as in the 2D case,
which is known by observing that the function
\eq
g(\bn, \bn', \bn) = -\f23 + 2 (\bn \cdot \bn')^2 
\qe
takes its minimum when $\bn \perp \bn'$.

The domain reconfiguration
due to the proper elastic interaction is
suppressed by the Frank elasticity 
at wavelengths shorter than
\eq
\xi_c = \sqrt{\f{K}{\mu \alpha^2}}.
\qe
Thus we have three characteristic lengthscales,
$\xi$, $\xi_R$, and $\xi_c$.
The observed domain size $\xi$ is typically
$1-10^1 \mu$m, while we estimate $\xi_c$ to be $10$ nm
for typical experimental values
$K=10^{-11}$ J/m, $\mu=10^5$ J/m$^3$, and $\alpha=1.0$.
There is a substantial gap between $\xi$ and $\xi_c$,
where the proper elastic interaction plays a dominant role.
The Frank free energy density $f_F$ (averaged over space)
scales as
$f_F \sim f_{el}^P \cdot (\xi_c/\xi)^2 \ll f_{el}^P\sim \mu \alpha^2$.
The domain size $\xi$
can be cast into a scaling form,
\eq
\f{\xi}{\xi_R} = \Xi\lrS{D, \f{\xi_c}{\xi_R}}.
\qe
Although $\Xi$ is a highly non-trivial function,
it can be numerically obtainable
unless $D$ is very small (or, unless $\xi/\xi_R$ is very large),
as we see in Section III.
We have a trial estimate $D\sim 1$
if we assume
$\beta \sim 0.01$ and
$\xi_R \sim 100$ nm in addition to the above values 
of $K$, $\mu$, and $\alpha$.
Of course, this estimate of $D$ is quite uncertain
because the magnitudes of $\beta$ and $\xi_R$
should depend on 
the kinetics of the crosslinking process, 
quality of the solvent, etc.
Our point here is that 
it is not unreasonable to have
a moderately strong disorder
in the presence of
submicron-scale network heterogeneities,
which is considered ubiquitous.

\subsection{Mechanical response}

Now we proceed to discuss the mechanical response
during the polydomain-monodomain transition.
To do so, it is useful to examine again 
the polydomain state at $\lambda=1$ and in 2D.
The harmonic free energy \Ref{Felproper} 
can be rewritten as~\cite{Uchida-Onuki}
\eq
\tilde{F}^P_{el} &=& -\f{\mu\alpha^2}{2} \int_{\bq} \biglrF{Q_1(\bq)}^2,
\label{Fel2D}
\\
Q_1(\bq) &=& 2 \qh_x \qh_y Q_{xx}(\bq) - (\qh_x^2 - \qh_y^2) Q_{xy}(\bq)
\nn\\
&=& \sin 2\varphi \; Q_{xx}(\bq) - \cos 2 \varphi \; Q_{xy}(\bq),
\label{Q1}
\qe
where
$\varphi$ is the azimuthal angle
of the wavevector, $\bq = |\bq|(\cos \varphi, \sin \varphi)$.
Complementary to $Q_1(\bq)$ is the variable 
defined by
\eq
Q_2(\bq) &=& (\qh_x^2 - \qh_y^2) Q_{xx}(\bq) + 2 \qh_x \qh_y Q_{xy}(\bq)
\nn\\
&=& \cos 2\varphi \; Q_{xx}(\bq) + \sin 2 \varphi \; Q_{xy}(\bq).
\qe
Note that $Q_1(\bq)$ and $Q_2(\bq)$ constitute
a set of normal modes, and satisfy
\eq
\biglrF{Q_1(\bq)}^2 + \biglrF{Q_2(\bq)}^2 
=
\biglrF{Q_{xx}(\bq)}^2 + \biglrF{Q_{xy}(\bq)}^2,
\label{normalmode}
\qe
or
\eq
\Av{Q_1(\br)^2} + \Av{Q_2(\br)^2} = 
\Av{Q_{xx}(\br)^2} + \Av{Q_{xy}(\br)^2} = \f14,
\label{normalmode2}
\qe
where $Q_a(\br)  \; (a=1,2)$ 
are the inverse Fourier transform of $Q_a(\bq)$.
To reduce the free energy (\ref{Fel2D}),
there arises an asymmetry  $Q_1(\br)^2 > Q_2(\br)^2$.
In the limit where $\mu \alpha^2$
is much larger than the disorder and Frank contributions
to the free energy density,
we expect from \Ref{normalmode2} to have
\eq
\Av{Q_1(\br)^2}=\f14,  \quad  \Av{Q_2(\br)^2}=0,  
\label{varQ1Q2}
\qe
which indeed is numerically confirmed~\cite{Uchida}.
In this limit, the elastic free energy density is given by
\eq
f_{el} = \f{\mu\alpha^2}{8},
\qe
as seen from (\ref{Fel2D}).
To the second order in $\alpha$,
it is equal to the free energy in the
monodomain state with $\lambda = \lambda_m$,
as we can easily check by
substituting (\ref{lambdam}) into
(\ref{Felbasic}) and expanding it with respect to $\alpha$.
Thus we conclude that the elastic free energy
change accompanied with the P-M
transition is of $O(\alpha^3)$,
and the macroscopic stress averaged over the region $1<\lambda<\lambda_m$, or
\eq
\f{f_{el}(\lambda=\lambda_m)-f_{el}(\lambda=1)}{\lambda_m-1},
\label{avstress}
\qe
is a quantity of $O(\alpha^2)$.

To see the origin of the soft response,
it is useful to look at 
the local elastic stress tensor, 
which is given 
in the harmonic approximation~\Ref{basicDu}
as~\cite{stress_tensor}
\eq
\sigma_{ij} = \mu 
\bigglrL{\f12 (\D_i u_j + \D_j u_i) - \alpha Q_{ij} + R_{ij}}.
\label{local_stress}
\qe
Consider
its variance $\Av{\sigma^2_{ij}}$.
In the absence of random stresses, we have
\eq
\int d\br \; \sigma_{ij}^2 &=& 
\mu F_{el} - \mu^2 \alpha^2 \int d\br \; Q_{ij}^2
\nn\\
&=& \mu^2 \alpha^2 \int d\br \biglrS{Q_1^2 - Q_{ij}^2}
\nn\\
&=& \mu^2 \alpha^2 \int d\br \; Q_2^2,
\qe
which vanishes from (\ref{varQ1Q2}).
This means that each part of the system is
stretched along the local director 
by $1+\alpha/4+O(\alpha^2) \sim \lambda_m$ times.
This local elongation, 
realized by the checkered polydomain structure,
reduces the free energy close to its absolute minimum.

\section{Numerical Simulation}

To further study non-linear mechanical
response and effect of random disorder,
we resort to numerical simulation by the continuum model.
We utilize two different numerical schemes,
one for the polydomain state in
mechanical equilibrium at $\lambda=1$
and another for the P-M transition and dynamical effects.
A two dimensional system is assumed
for computational advantage.
All the simulations below are
performed on a $N \times N$ square lattice
with $N=128$ unless otherwise stated.
The grid spacing is
chosen to be the unit of length.
Periodic boundary conditions
are imposed on $\bn(\br)$ and $\bu(\br)$,
while the average strain $\lambda$ is externally 
controlled.

\subsection{Polydomain state}

First we study the the polydomain state
in complete mechanical equilibrium
and with no average strain ($\lambda=1$).
To this end, we assume the harmonic 
free energy \Ref{basicDu} and solve 
the linear equations \Ref{equilibrium} and \Ref{incompressibility}
using fast Fourier 
transform. 
To minimize the free energy,
we adapted a variant of the simulated 
annealing method~\cite{NumericalRecipe}.
The orientational order parameter is evolved
according to a Langevin equation,
\newcommand{\bmeta}{\bm \eta}
\eq
\der{\bn}{t} &=& \Gamma_n \,(\tI-\bn\bn)
\cdot \bigglrS{- \der{F}{\bn} + \bmeta},
\label{Langevin}
\qe
where $\Gamma_n$ is a constant and
$\bmeta$ is a ``thermal'' noise
satisfying
\eq
\lrA{\bmeta(\br,t) \bmeta(\br',t')} = \eta_0^2 \; \tI \cdot \delta (\br-\br')
\delta(t-t')
\qe
and Gaussian statistics.
The noise strength $\eta_0$
is gradually reduced to zero until the end of each run.
To be precise, we decrease $\eta_0$ to zero
at a constant rate in the former half of a run,
and set $\eta_0=0$ in the latter half.
The initial noise strength and the annealing rate
are chosen so that two different
initial configurations, 
one with random and 
another with homogeneous director field, 
lead to indistinguishable 
results for the macroscopic quantities
such as correlation function,
average orientation, and free energy densities.
As a standard set of static parameters we choose
\eq
\mu=400, \, 
\alpha=0.2, \, 
\beta =0.025, \,
\xi_R = 1,
K=4, 
\label{standard_parameter}
\qe
for which $\xi_c=0.5$ and $D=0.5$.
We integrated \Eq{Langevin} using the Euler scheme 
with time increment $\Delta t=1$ per step.
A typical run consisted of $5\times10^4$ time steps.
Longer runs did not make an  observable difference
in the macroscopic quantities.

First we consider the orientational correlation function,
\eq
G(\bR) &=& \bigglrA{ Q_{ij}(\br) Q_{ij}(\br + \bR)},
\qe
which is a function only of distance.
We define the correlation length $\xi$ through
\eq
\f{G(\xi)}{G(0)} = \f12.
\qe
For each parameter set, we took
statistical average over 20 samples.
The data are shown in \Fig{corrG}.
The decay of $G(R)$ is nearly exponential
for strong disorder and faster than exponential
for weak disorder.
This qualitative tendency agrees with
previous results for the 2D random-field XY
model~\cite{Dieny-Barbara,Gingras-Huse,Yu-etal}.
The correlation length is a rapidly 
decreasing function of
the effective disorder strength, $D$.
The dependence is roughly exponential,
also in agreement with previous results
for the XY model~\cite{Dieny-Barbara,Yu-etal}.
In the same figure 
we show the dependence of $\xi$
on $\mu\alpha^2$, which is the measure of elastic interaction.
Although the dependence is weak,
the proper elastic interaction has an effect of increasing
the correlation length.
This is related to
the enhancement of correlation in 
directions oblique to the local director, 
depicted in \Fig{preferN2D}.
In order to quantify the director-relative correlation,
we define the function
\eq
H(\bR) &=& \bigglrA{ Q_{ij}(\br) \, Q_{ij}(\br + \tensor{U}(\br) \cdot \bR)},
\qe
where $\tensor{U}(\br)$ is a matrix of rotation 
that maps $\bn(\br)$ to $\be_x$, 
or, explicitly,
\eq
\tensor{U} &=& \left[ 
\begin{array}{cr} 
\cos \theta & -\sin \theta \\
\sin \theta &  \cos \theta
\end{array} \right].
\label{relativecorr}
\qe
By definition,
$H(x, 0)$ and $H(0, y)$ respectively describe 
the correlation in directions parallel and perpendicular 
to the local director.
The data for the standard parameter
are plotted in \Fig{corrT}.
We see that the correlation is long-ranged
in any specific direction,
and the exponential-like decay in \Fig{corrG}
should be considered as a result of 
mutual cancellation of positive and negative
correlation by taking the angular average.

A real space snapshot of
the order parameter field $Q_{xy}$
is also given in \Fig{corrT}.
As the grayscale shows,
the contour $Q_{xy}=0$
preferentially lies in the
horizontal ($x$-) and vertical ($y$-) directions.
This corresponds to the checkered domain structure 
in \Fig{preferN2D}
(note that the grayscale is chosen
so that the director is oblique to the horizontal axis
in the brightest and darkest regions).
More precisely,
the checkered pattern is
found on many different lengthscales,
which is a natural consequence
of the fact that the elastic interaction energy \Ref{FeldD}
is scale-independent.

An experimentally accessible way to characterize
the anisotropic director correlation is
the polarized light scattering.
In a weakly inhomogeneous state,
the depolarized (HV) light scattering intensity
is given by
\eq
I(\bq) = \bigglrA{\biglrF{Q_{xy}(\bq)}^2},
\label{Iqformula}
\qe
except for a $\bq$-independent prefactor.
According to
Ref.~\cite{Clarke2},
the above formula holds
even in a highly inhomogeneous state,
if one assumes a two-dimensional  configuration
(see Eq.(2) in the reference).
Our numerical data is shown in \Fig{corrT}.
The intensity \Ref{Iqformula} is expressed in terms of $Q_1$ and $Q_2$ as
$I(q)=\cos\varphi^2\lrA{|Q_1(\bq)|^2}+\sin\varphi^2\lrA{|Q_2(\bq)|^2}$,
and the asymmetry $Q_1 > Q_2$ explains the enhanced
scattering on $q_x$- and $q_y$- axes~\cite{Uchida-Onuki}.

Note that the peak is located at a small but finite wavenumber,
in contrary to what is expected from
the non-conserved nature of the
orientational order parameter.
In fact, we find it to be a finite size effect, and
the peak wavenumber shrinks to zero as the system size $N$
is taken to infinity, leaving a singular minimum at the origin.
To see this,
we have computed the circularly averaged 
structure factor,
\eq
S(q) = \int_{0}^{2\pi} \!\!d\varphi\; \bigglrA{\biglrF{Q_{ij}(\bq)}^2},
\label{defSq}
\qe
for $N=64$,$128$, and $256$ systems, and found a
peak in the region $(2\pi/N) < q < 2 (2\pi/N)$ in every case.
The origin of the singular minimum at $q=0$ is
explained as follows.
Because of the periodic boundary condition
on $\bu$, the spatial average $\Av{\grad \bu}$
should complete vanish.
This constraint suppresses
formation of the checkered pattern
with the check size larger than $N/2$.

\subsection{P-M transition}

Next we study the P-M transition
using the non-linear elastic free energy \Ref{Felbasic}.
We found that complete minimization of the
free energy takes very much computation time,
and decided to take a more empirical approach : 
we utilize a simple dynamical model,
and abandon to exclude non-equilibrium effects
from the results.
Fortunately, the stress-strain relation thus
obtained is equilibrated to a good degree,
because of fast relaxation of the rubber-elastic
free energy.
On the other hand, the domain structure
exhibits a slow coarsening, which we study
in the absence of external strain.

Our dynamical model consists of 
a set of equations that describe
evolution of
non-conserved order parameters in a simplest manner,
namely,
\eq
\der{\bn}{t} &=& - \Gamma_n \,(\tI-\bn\bn) \cdot \der{F}{\bn},
\label{relaxQ}
\qe
and
\eq
\der{\bu}{t} &=& - \Gamma_{u} \, \fder{F}{\bu}.
\label{relaxU}
\qe
Instead of imposing the strict
incompressibility condition \Ref{incompressibility},
we penalized local volume change 
by adding an artificial potential $F_v$
to the free energy.
By taking it in the form $F_v = \f12 \int d\br 
\lrL{ a_0 (\det \tensor{\Lambda}-1)^2
+     a_1 (\det \tensor{\Lambda}-1)^4}$
and choosing appropriate values of the constants $a_0$ and $a_1$,
we kept $\det \tensor{\Lambda}$
in the region $[0.99,1.01]$ throughout the runs.

We integrated \Ref{relaxQ} and \Ref{relaxU} using the Euler scheme
with $\Gamma_{n}=0.2$, $\Gamma_{u}=0.02$ and $\Delta t =1$.
A typical set of static parameters is
same to that given by \Ref{standard_parameter}.

To prepare a polydomain state,
we set site-wise random numbers to $\tQ$ and $\bu$
as the initial condition,
and integrated \Ref{relaxQ} and \Ref{relaxU} for
$5\times 10^4$ time steps with $\lambda=1$.
Then we increased $\lambda$ at a constant
rate $d\lambda/d t =1\times10^{-5}$
to induce the P-M transition.
To check hysteresis, finally we decreased $\lambda$
back to unity at the rate
$d\lambda/d t = -1\times10^{-6}$.

Plotted in \Fig{stress_orientation}
are the scaled macroscopic elastic 
stress $\mu^{-1} \sigma_{macro}= \mu^{-1}(\D f_{el}/\D \lambda)$
and the mean orientation $S = \cos 2\theta = 2Q_{xx}$
as functions of $\lambda$.
We see from the figure that the
elastic stress is vanishingly small
and the orientation linearly increases
in the polydomain region $1<\lambda<\lambda_m(=1.05)$.
The stress shows a linear rise
in the monodomain region $\lambda > \lambda_m$,
where the orientation is nearly saturated to the maximum, $S=1$.
While the strain-orientation curve has a small hysteresis,
the strain-stress curve is almost completely reversible.

The smallness of the hysteresis manifests an important
difference between the present system and 
random anisotropy magnets under magnetic 
field.
In the latter, the macroscopic orientational order
is broken solely by a random field.
In contrast, in the present model,
the monodomain state is unstable
to an strain-mediated director buckling
for $\lambda<\lambda_m$,
even when there is no quenched disorder.
This instability, which will be discussed in Section IV
in detail, makes the P-M transition almost reversible.

The free energy densities
are also shown in \Fig{stress_orientation}.
Both the proper and disorder
parts of the rubber-elastic free energy
change little in the region $\lambda<\lambda_m$.
The latter curve has a slightly positive gradient.
The situation is more subtle for the former.
Its gradient is slightly positive in the figure,
and turns to slightly negative for a smaller
disorder strength.
However,
in the absence of random
stresses and at $\lambda=1$, 
we had four domains whose sizes are
limited by the system size, and the domain boundaries
raise the elastic free energy. Because of this finite 
size effect, 
we cannot exactly tell
the sign of the proper elastic stress
in the macroscopic limit.
We cannot exclude the possibility that
the macroscopic stress completely vanishes in the
limit of weak disorder.

The strain-stress and strain-orientation curves for larger values of
$\alpha$ are given in \Fig{stress_orientation_alpha}.
Each curve shows a sharp crossover
around $\lambda=\lambda_m(\alpha)$.
The elastic stress in the polydomain region
is vanishingly small even for large coupling.
For any value of $\alpha$ studied,
the changes of the proper and disorder elastic
free energy densities,
$|f^P_{el}(\lambda=\lambda_m)-f^P_{el}(\lambda=1)|$
and
$|f^D_{el}(\lambda=\lambda_m)-f^D_{el}(\lambda=1)|$,
were smaller than
$0.3$ percent of $\mu\alpha^2$.
We find essentially no $\alpha$-dependence
of the macroscopic stress in the polydomain region.

Shown in \Fig{histogram} is the histogram of
the elastic free energy contained in a lattice site.
The distribution is fairly sharp and
little changed by stretching
for $\lambda<\lambda_m$,
implying that the free energy 
is homogeneously minimized 
in the polydomain state.
Real space snapshots of the domain
morphology is shown in \Fig{domain_morphology}.
Pinned defects are observed
just below the threshold $\lambda=\lambda_m$,
while we find no defects remaining
in the monodomain state.

The depolarized  scattering 
intensity is shown in
\Fig{scattering_intensity}.
It has a minimum at $q=0$ and develops four peaks
at finite wavenumbers.
As we shall see in the next subsection,
the peaks move toward the origin
as the true equilibrium is approached 
and the domains coarsen.
Here we concentrate on the effect of stretching.
The peak intensity first increases
and then decreases as a function of $\lambda$.
Under stretching along the $x$-axis,
the peaks on the $q_x$- axis
are more enhanced than those on the $q_y$- axis.
The shift of peak wavenumber 
by stretching is very small
and difficult to estimate.
By our choice of the stretching rate,
the P-M transition completed in 
$5\times 10^3$ time steps, much before a 
significant coarsening can occur.

\subsection{Slow structural relaxation}

Now we turn to dynamical effects.
First let us discuss
the conditions under which
Eqs.(\ref{relaxQ}) and (\ref{relaxU})
are most reasonable
as a model of dynamic evolution,
not only as an artificial scheme of
functional minimization.
Firstly, \Eq{relaxU} means
that the velocity $\D\bu/\D t $
is proportional to the force $-\delta F/\delta\bu$.
This applies to motion of a network
in a viscous solvent~\cite{TwoFluid},
where we have a straightforward analogy to D'arcy's law 
in porous media.
On the other hand, in dry elastomers,
there arises a viscous stress due to
intra-network friction, which is
proportional to $\grad (\D\bu/\D t)$.
This is not accounted for in \Eq{relaxU}.
Thus we consider that the dynamic model is
more appropriate to swollen gels
than to elastomers.
Secondly, Eqs.(\ref{relaxQ}) and (\ref{relaxU})
neglect dynamical coupling between the order parameters,
i.e., the non-diagonal part of the Onsager coefficient matrix.
This does not matter if the dynamics of the orientational
order parameter is fast and slaved to that of the
displacement field, which we expect to be the case.
In fact, if the constituent polymer of the gel is not rigid,
$\Gamma_n^{-1}$ is of the order of the viscosity of
low-molecular weight fluids, $\eta$.
On the other hand, the friction between the
network and solvent renders $\Gamma_u^{-1}$ to be of the
order of $\eta/l^2$, where $l$ is the mesh size
of the network~\cite{OnukiGel,Harden-etal}. 
Thus, the characteristic relaxation time of 
the strain at the scale of domain size 
is $\sim (\xi/l)^2 (\gg 1)$ 
times larger than that of $\tQ$.

Evolution of the structure factor $S(q)$ (as
defined by \Eq{defSq})
is shown in \Fig{relaxationSq}.
The peak wavenumber decreases
and the peak intensity increases
as a function of time.
Also shown in the figure is
the structure factor at
complete mechanical equilibrium,
which is
obtained by the numerical scheme used in Section III A.
The correlation length $\xi$
and the inverse of the peak wavenumber $q_0$
are plotted in the middle of \Fig{relaxationSq}.
In the time region $1\times 10^3 < t < 3\times10^5$,
the former is well fitted by a power law $\xi(t) \propto t^{\epsilon}$
with $\epsilon =  0.23 \pm 0.02$, and the latter
grows almost in parallel to the former.

The Frank and rubber-elastic free energies
are plotted as functions of time
in \Fig{relaxationEnergy}.
While the elastic free energy
changes little after
an early stage of around $t=10^3$ time steps,
the Frank energy density $f_{F}$
shows a slow and continuous decrease, which is approximately described
by an power law $f_{F}\propto t^{-\epsilon'}$
in the region $1\times10^3<t<3\times10^5$,
with $\epsilon'=0.22\pm0.03$.

Presently we have no explanation for
the good fits of $\xi(t)$ and $f_F(t)$
by power laws.
We keep ourselves to point out that
the values of $\epsilon$ and $\epsilon'$
are much smaller than the corresponding
exponents for the 2D non-conserved XY model 
without quenched disorder,
which equal $0.5$ and $1.0$
from a simple scaling argument~\cite{Bray_review}.
The naive scaling relation $\epsilon' = 2 \epsilon$
is also broken here, which is not at all surprising
if we consider the presence of 
quenched disorder~\cite{Zapotocky-etal}.
We should also stress that
the final equilibrium values of $\xi$ and $f_{F}$ are finite.
Preliminary study by a longer run
without statistics finds 
a crossover from the power-law type
kinetics to a slower one
at $t\sim 1\times10^6$ steps.

The above results show that
the relaxation process
can be decomposed 
into three characteristic stages :
(i) The quench into the nematic phase
from the isotropic phase
produces microscopic textures,
which coarsen to reduce both
the rubber-elastic and Frank free energies.
After the characteristic domain size
reaches $\xi_c$,
anisotropic domain reconfiguration
on this scale follows.
The rubber-elastic free energy
is almost completely minimized at this early stage,
because of the scale-independence
of the proper elastic interaction \Ref{FeldD}.
(ii) The ``checkered'' domain structure
further coarsens to reduce the Frank free energy.
The domain size $\xi$ and the peak wavelength $2\pi/q_0$
grow in parallel to each other.
(iii)
The domain size converges to a finite
equilibrium value,
while the anisotropic domain reconfiguration
proceeds on larger scales ($\lim_{t\to \infty}q_0(t) = 0$).

\section{Fluctuation in the monodomain state}

Recall that, if there is no quenched disorder,
the ground state of the system is the macroscopically
elongated state with $\lambda=\lambda_m$ and $\bu=0$.
In this state, there are so-called soft modes of
director fluctuation, which do not accompany
any change in the rubber elastic free
energy~\cite{Golubovic-Lubensky,WarnerSoft,Olmsted}.
The presence of the soft modes implies
that a homogeneous director configuration
becomes unstable
for $\lambda<\lambda_m$;
when we compress
the gel along the optical axis,
the director ``buckles'' to
partially cancel the rise of elastic
free energy by compression.
The result of the previous section means
that this instability is almost completely
soft even for large deformations.
In this section, we look at the monodomain
region $\lambda\ge\lambda_m$
and analyze the director fluctuation modes in a harmonic level.
It was suggested in Ref.~\cite{Golubovic-Lubensky}
that the soft fluctuations at the critical point $\lambda=\lambda_m$
is strongly enhanced by quenched disorder
and satisfy $\lrA{|\delta\bn(\bq)|^2} \propto q^{-4}$.
Nonetheless, it was not
fully confirmed because of a breakdown of the harmonic approximation
at the critical point, $\lambda=\lambda_m$.
Also, the model used in~\cite{Golubovic-Lubensky}
remains largely phenomenological, and a quantitative 
assessment of the prediction is necessary.
Here we extend the analysis to arbitrary values 
of $\lambda$, and discuss the possibility to
find disorder-enhanced fluctuations in 
practical situations.

The director is decomposed into a homogeneous part and a small deviation,
as
\eq
\bn(\br) = \be_x + \delta \bn(\br).
\label{deltan}
\qe
Expanding the basic free energy \Ref{Felbasic}
with respect to $\delta \bn$ and $\bu$,
and then eliminating the elastic field 
using the mechanical equilibrium condition,
we obtain
an effective free energy in terms of $\delta \bn$.
An outline of the calculation is given in Appendix A.
For the three dimensional case,
the result is 
\eq
\tilde{F}_{el} &=& 
\f{\mu \alpha}{\lambda} 
\int_{\bq} \biggl[
\f12 A_1(\lambda, \bqh) \,  \biglrF{\delta \bn (\bq)}^2
+ \f12 A_2(\lambda,\bqh) \, \biglrF{\bqh \cdot \delta \bn (\bq)}^2
\nn\\ 
&&\hspace{15mm}
- R'_{ix}(\bq) \delta n_i (-\bq) 
\nn\\ 
&&\hspace{15mm}
- B_1(\lambda,\bqh) \qh_i R'_{ix} (\bq)
\bigglrS{ \bqh \cdot \delta \bn (-\bq)} 
\nn\\ 
&&\hspace{15mm}
- B_2(\lambda,\bqh) 
\bigglrS{ \bqh \cdot \tR'(\bq) \cdot \delta \bn(-\bq)} 
\nn\\ 
&&\hspace{15mm}
- B_3(\lambda,\bqh) 
\bigglrS{\tR'(\bq) : \bqh \bqh} 
\bigglrS{\bqh \cdot \delta \bn (-\bq)} 
\biggr],
\label{Feffmono}
\qe
\eq
A_1(\lambda, \bqh) &=& 
\lambda^3 - 1  - 
\f{3 \alpha}{3 +4\alpha}\f{\lambda^6 \qh_x^2}{1 + (\lambda^3-1)\qh_x^2},
\\
A_2(\lambda,\bqh) &=& 
\f{ 
\f{3 \alpha (3+4\alpha)}{3+\alpha}
\f{(\f{3+\alpha}{3-2\alpha}+\lambda^3)^2 \qh_x^2}{3+\alpha(2+\qh_x^2)}
-\f{3\alpha}{3-2\alpha}
}
{1 + (\lambda^3-1)\qh_x^2},
\\
B_1(\lambda, \bqh) &=& \f{1+\f{3\alpha}{3+\alpha}
(\f{3+\alpha}{3-2\alpha}+\lambda^3)\qh_x^2}{1+(\lambda^3-1)\qh_x^2},
\\
B_2(\lambda, \bqh) &=& \f{\lambda^3 \qh_x}{1+(\lambda^3-1)\qh_x^2},
\\
B_3(\lambda, \bqh) &=& -\f{(\f{3+\alpha}{3-2\alpha}
+\lambda^3)\qh_x}{1+(\lambda^3-1)\qh_x^2},
\\
R'_{ij}(\bq) &=& \Av{\Lambda}_{ik}\Av{\Lambda}_{jl} R_{kl}(\bq),
\qe
which is correct to the bilinear order
in $\delta\bn$ and $\tR$ (we neglect
terms independent of $\delta \bn$).

In the absence of quenched disorder,
the integrand in \Ref{Feffmono} is of the form
$\f12 (A_1 \tI + A_2 \bqh \bqh)  : \delta \bn(\bq) \delta \bn(-\bq)$.
It is convenient to
introduce
two unit vectors
$\be_1=\bq\times \be_x/|\bq\times \be_x|$
and $\be_2= \be_1 \times \be_x$~\cite{deGennesLCbook},
with which the integrand becomes
\eq
\f12 A_1 \biglrF{\be_1 \cdot \delta \bn(\bq)}^2 
+ \f12 \bigglrS{A_1 + A_2(1-\qh_x^2)} \biglrF{\be_2 \cdot \delta \bn(\bq)}^2.
\qe
At $\lambda=\lambda_m$,
the coefficient $A_1$
takes its minimum value $0$ for $\bq \parallel \be_x$,
while $A_1+A_2(1-\qh_x^2)$ is minimized and vanishes
both on the line $\bq \parallel \be_x$ 
and in the plane $\bq \perp \be_x$.
These correspond to the soft modes.
Similar results have been obtained
by Olmsted~\cite{Olmsted}
for monodomain elastomers
crosslinked in the nematic phase
and without external strain.

The random stresses shift the ground state
to an inhomogeneous state, 
$\delta \bn = \delta \bn_R$ and $\bu=\bu_R$.
The frozen director deviation $\delta \bn_R$
is obtained by
minimizing the total free energy $\tilde{F}_{el}+F_F$
with respect to $\delta \bn$, as
\eq
\delta n_{R,i}(\bq) &=& \f{1}{A_1 + K' q^2}
\bigglrL{
R'_{ix}(\bq) + B_2 \; \qh_j R'_{ij}(\bq)
\nn\\&&
+ \bigglrS{B_1 - \f{A_2(1+B_1)}{A_1+A_2 + K' q^2}} 
\qh_i \qh_j R'_{jx}(\bq)
\nn\\&&
+ \bigglrS{B_3 - \f{A_2(B_2+B_3)}{A_1+A_2+K' q^2}}
\qh_i \bigglrS{\bqh \bqh : \tR'(\bq)}
},
\label{deltanR}
\qe
where we have introduced a scaled Frank constant,
\eq
K' &=& \f{K \lambda}{\mu \alpha}. 
\qe
At the critical point $\lambda = \lambda_m$,
the quantity $A_1 + A_2$ appearing in \Ref{deltanR}
vanishes for $\bq\perp \be_x$.
Hence, in the long wavelength limit $q\to0$,
we have $\delta n_R (\bq) \propto q^{-2}$
in the soft directions $\bq\perp\be_x$ and $\bq \parallel \be_x$.
This means a divergence of the real space
amplitude $\lrA{|\bn_R(\br)|^2}$
and breakdown of the harmonic approximation,
as pointed out in Ref.\cite{Golubovic-Lubensky}.
Severer is the divergence of
the frozen elastic field $\bu_R$, which
is related to $\delta \bn_R$ through
the mechanical equilibrium condition \Ref{equilibrium2}.
At $\lambda=\lambda_m$,
it behaves as $\lrA{|\bu_R(\bq)|^2} \propto q^{-6}$
in the soft directions, and $\lrA{|\bu_R(\br)|^2}$
diverges.
However, these divergences disappear
for $\lambda>\lambda_m$,
where the excess stretching
acts as a stabilizing field.
Now we consider this region.
The condition for the harmonic approximation to be valid
is $\lrA{|\delta \bn_R(\br)|^2} \ll 1$,
which implies $\lrA{|\grad \bu_R(\br)|^2} \ll \alpha^2$.
To assess the condition by order estimate,
we concentrate on the plane $\bq\perp\be_x$,
from where arises the most significant contribution to
the real space amplitude, 
\eq
\lrA{\lrF{\delta \bn_R(\br)}^2} = \f{1}{(\mbox{system's volume})}
\int_{\bq} \lrA{\lrF{\delta \bn_R(\bq)}^2}. 
\qe
On that plane, the strongest $q$-dependence of $\delta\bn_R(\bq)$  
comes from a factor $(\Delta + K'q^2)^{-1}$, where
\eq
\Delta = \Delta(\lambda) = 
\biglrL{A_1(\lambda,\bqh)+ 
A_2(\lambda,\bqh)}\big|_{\qh_x=0}.
\qe
is the measure of excess stretching. Note that
$\Delta \propto \lambda-\lambda_m$ for $\lambda-\lambda_m\ll 1$.
A saddle-point approximation around the plane
yields
\eq
\lrA{\lrF{\delta \bn_R(\br)}^2} &\sim& \int 
\f{\beta^2 \xi_R^3 q^2 dq}{(\Delta + K'q^2)^{3/2}}
\nn\\
&& \hspace{-20mm}
\left\{
\begin{array}{cc} 
\displaystyle
\sim \f{\beta^2 \xi_R^3}{K'^{3/2}}
\ln\bigglrS{\f{K' q_{max}^2}{\Delta}} 
\quad & (\Delta \ll K' q_{max}^2),
\\
\displaystyle
\sim \f{\beta^2 \xi_R^3}{K'^{3/2}}
\bigglrS{\f{K'q_{max}^2}{\Delta}}^{3/2} 
\quad & (\Delta \gg K' q_{max}^2),
\end{array}
\right.
\label{deltanR2}
\qe
where $q_{max}$ is the upper cutoff wavenumber.
We may roughly identify $2\pi/q_{max}$ with the network mesh size,
$l \sim (k_BT/\mu)^{1/3}$.
Using typical experimental values $\mu=10^5$ J/m$^3$, $K=10^{-11}$ J/m,
$\alpha=1.0$ and $T=300$ K,
we have $\sqrt{K'}=10$ nm and $K'q_{max}^2\sim 1$.
For $\Delta \simle K' q_{max}^2$
and except in the close vicinity of
the criticality, $\Delta=0$,
the amplitude only weakly depends on the elongation ratio.
In this region,
the harmonic approximation is valid if and only if
\eq
\f{\beta^2 \xi_R^3}{K'^{3/2}} \ll 1,
\label{smallfluctuation}
\qe
which is satisfied when we put
$\xi_R\sim 10^2$ nm and $\beta\sim 0.01$ as a trial.

The director exhibits a thermal fluctuation 
around the inhomogeneous ground state.
Its amplitude is not affected by the quenched 
randomness, at least within the harmonic calculation.
The total fluctuation amplitude  
is given by
\eq
P^{(a)}(\bq) &=& 
\biglrA{
\biglrF{\be_a \cdot \delta \bn(\bq)}^2
}
= P^{(a)}_{T}(\bq) + P^{(a)}_{R}(\bq),
\\
P^{(a)}_{T}(\bq) &=& \f{k_BT\lambda}{\mu \alpha}
\f{1}{A_1 + \delta_{a2} (1-\qh_x^2) A_2 + K' q^2},
\\
P^{(a)}_{R}(\bq) &=& \biglrA{\biglrF{\be_a \cdot \delta \bn_R(\bq)}^2},
\qe
where $a=1,2$, and $P_T^{(a)}$ and $P_R^{(a)}$
are the thermal and frozen contributions,
respectively.
Let us compare the two contributions.
To be explicit, we compare $P_T^{(2)}(\bq)$
and $P_R^{(2)}(\bq)$ on the plane $\bq \perp \be_x$.
There the ratio $P_R^{(2)}/P_T^{(2)}$
is controlled by a factor of the form
$\Delta_c/(\Delta +K'q^2)$,
where 
\eq
\Delta_c = \f{\mu \alpha \beta^2 \xi_R^3}{k_BT}
\qe
defines a crossover point.
In the region $\Delta \simle \Delta_c$,
the disorder part of the fluctuation
dominates the thermal part
at long wavelengths.
We estimate $\Delta_c \sim 10^{-3}$
using the above mentioned values,
for which the crossover length 
$\sqrt{K'/\Delta_c} \sim 10^{2}-10^{3}$ nm
is around or below the wavelength of visible light.
The amplitudes for $\alpha=0.2$ and $\lambda/\lambda_m=1.001$ 
are plotted in \Fig{ozdirt}. We see that
the anisotropy of the scattering pattern
is not much affected by the quenched disorder.

The director fluctuation amplitudes
are closely related to polarized
light scattering intensity~\cite{deGennesLCbook}.
By comparing 
experimental results
to the above calculation,
we may extract information
on the network heterogeneity.
In particular, 
if the macroscopic orientation is
not saturated in the monodomain state,
it means the presence of large-scale 
quenched strains that do not meet
the condition \Ref{smallfluctuation}.
For instance, in an optical study of a swollen 
monodomain gel
by Chang et al.~\cite{Chang-etal},
speckles on the few-micrometer scale
are observed, which is attributed to
heterogeneities as we consider here.
We hope that the scattering intensity will be
measured as a function of applied strain.
Another origin of large scale heterogeneity
will be discussed in the next section.

\section{Effect of crosslinking condition}

In this section, we consider the
case of anisotropic crosslinking.
Melts of nematic polymers
often exhibit long-lived polydomain
textures after a quench from the isotropic 
phase~\cite{LCPpolydomain1,LCPpolydomain2,Doi-EdwardsBook}.
The size of the domains is macroscopic 
and typically of micron order.
When such a melt is crosslinked,
its non-uniform orientation is imprinted 
into the network.
We denote the initial configuration by $\tQ_0(\br)$.
The extended affine-deformation theory
prescribes the elastic free energy,
\eq
F_{el} = \f{\mu}{2} \int d\br \;
\Tr \bigglrL{
(\tI + \alpha_0 \tQ_0) \cdot \Lambda^T \cdot (\tI - \alpha \tQ) \cdot \Lambda
-\tI
}
\label{Felnematic},
\qe
where $\alpha_0$ is expressed in terms of the parameters
used in~\cite{Warner-TerentjevReview} as
\eq
\alpha_0 = 
\f{\ell_\parallel - \ell_\perp}{(1/d)\ell_\parallel + (1-1/d)\ell_\perp},
\label{alpha_0}
\qe
or equivalently, $\alpha_0  = \alpha / [1-(1-2/d)\alpha]$.
Note that the above free energy is
obtained by just formally replacing $\tR$ with $\alpha_0 \tQ_0$
in the free energy \Ref{Felbasic}.
Thus, the initial texture field $\tQ_0$
provides a source of quenched disorder.
An effective disorder strength that
corresponds to \Ref{disorder_strength} can be defined by 
\eq
D = \f{\mu\alpha\alpha_0}{K} \cdot \xi_0^2.
\qe
where  $\xi_0$ is
the correlation length of the initial texture.
If we set $\xi_0= 1 \mu$m and $\alpha=0.1$,
we have a very large number
$D \sim (\xi_0/\xi_c)^2 = 10^3 \sim 10^5$.
Since the orientational order
is predominantly affected by 
this strong disorder,
we do not take into account
other mesoscopic sources of quenched stresses, 
which is legitimate as a first approximation.

We have simulated the P-M transition 
in the following way.
To generate the initial configuration,
we mimicked the phase ordering kinetics of 
nematic polymer solutions by numerically solving the
equation,
\eq
\der{\bn}{t} &=& - \Gamma_n \,(\tI-\bn\bn) \cdot \der{F_F}{\bn}.
\label{modelA}
\qe
Taking a sitewise-random director
configuration as the initial condition,
we integrated \Eq{modelA}
over 50 time steps (with $\Gamma_n=0.2$ and $\Delta t = 1$)
to generate $\tQ_0(\br)$.
After crosslinking the system by adding $F_{el}$ to the free energy
and setting $\lambda=1$ and $\bu=0$,
we integrated Eqs.(\ref{relaxQ}) and (\ref{relaxU}) 
for $1\times10^4$ time steps
to equilibrate the system.
The mechanical response was studied 
in just the same way as described in Section III.

\Fig{stress_orientation_alpha_aniso}
shows the strain-stress and strain-orientation curves
for $\alpha=0.2,0.4,0.8,$ and $1.2$.
The strain-stress curve bends at a value of $\lambda$
where the director is yet far from aligned.
For $\alpha\ge0.8$,
the gradient of the strain-stress curve
has a non-monotonic dependence on the strain,
and is smallest at an intermediate value of $\lambda$.
The strain-orientation curve shows
only
a gradual crossover to the monodomain state,
especially for larger values of $\alpha$.
The slope of the scaled elastic stress $\mu^{-1}\sigma_{macro}$
is roughly independent of $\alpha$
in the vicinity of the point $\lambda=1$.

Evolution of the director texture
during the P-M transition
is shown in \Fig{domain_morphology_aniso},
and the distribution of the elastic
free energy in~\Fig{histogram_aniso}.
At $\lambda=1$ the director texture is almost
same as that just before crosslinking, or
$\tQ(\br)=\tQ_0(\br)$.
This is reflected in the
extremely homogeneous free energy
distribution.
External strain strongly dehomogenize
the distribution,
and the peak is continuously broadened
as we increase $\lambda$ toward the monodomain region.

\section{Discussion and Summary}

We have studied polydomain
nematic networks from two aspects, namely,
(i) breaking of long-range orientational order by frozen
internal stresses and
(ii) a non-local inter-domain interaction
arising from strain-orientation coupling.
The mechanical response is controlled by the latter
if the quenched disorder is of moderate strength (or, if $D\simle1$).
In this case, the proper elastic interaction reorganizes the polydomain 
structure so that local elongation along 
the director is achieved everywhere in the system.
The resulting structure contains
the checkered correlation pattern 
on various scales, which produces
the ``four-leaf clover'' pattern
in the depolarized scattering intensity.
Upon stretching, the director and
the local strain axis coincidently
rotate toward the direction of 
macroscopic extension,
and thus the elastic free energy
keeps almost constant until a complete 
alignment is attained.
The change in elastic free energy 
accompanying the P-M transition 
is analytically estimated to be of $O(\alpha^3)$.
This result does not depend on a specific model of 
non-linear elasticity. 
In fact, we obtained it by harmonic expansion 
of the elastic free energy, 
which is unique from symmetry~\cite{deGennesCRAS}.
Numerical simulation reveals a more complete softness,
and we find essentially no $\alpha$-dependence
of the average macroscopic stress \Ref{avstress}.
We cannot exclude the theoretical possibility that
the P-M transition is exactly soft 
in the weak disorder limit.
We may say that there are
mechanical {\it quasi-Goldstone modes},
which are distinguished from the genuine Goldstone modes
of fluid nematic liquid crystals
in that there is an anisotropic correlation
even in the absence of external field 
and that the quenched disorder selects a characteristic 
lengthscale.

We have discussed two sources of
quenched disorder.
One is the random stress due to
residual heterogeneous strains at the moment of crosslinking,
considered ubiquitous in rubbery networks.
The anisotropic (shear) part of random stresses
act on
the orientational order, both locally and non-locally.
The macroscopic domain size
observed in experiments can be explained
if there are frozen heterogeneities
of a reasonably small magnitude (e.g. $1$ percent in strain)
and a size somewhat
larger than that of individual network meshes
(e.g. $10^2$ nm).

A different viewpoint is taken in
previous theories~\cite{tenBosch-Varichon,Clarke1,Fridrikh-Terentjev},
where a random molecular field operating at crosslinks
is assumed to be the source of disorder.
The random field hypothesized there
has a small correlation length
roughly equal to the distance $l$ between crosslinks.
The ratio $\xi/l$ is a large number ($\sim 10^3$),
which means a weak effective disorder.

Currently we know of no firm experimental indication
of the disorder strength.
A possible method of its estimate is to observe
director fluctuation in the monodomain state.
We have calculated thermal and disorder contributions
to the fluctuation amplitude, which is proportional 
to the polarized light scattering intensity.
The intensity diminishes as we stretch the network,
and there is a region of
macroscopic strain $\lambda$
where the disorder contribution
dominates over the thermal one.
The width of the region and
the absolute value of the intensity
should inform us the order of the disorder strength.
The thermal and quenched contributions
could be separately analyzed by use of dynamical 
light scattering (DLS). Indeed, DLS has been 
successfully used to decompose
the two kinds of density fluctuation in gels~\cite{Shibayama}.
It is hoped that a similar method will be
developed for orientation 
fluctuation in the present system.

By crosslinking the network
in the nematic phase and
in the course of phase ordering,
we obtain another kind of quenched stresses.
The polydomain texture of the liquid-crystalline 
polymer melt is almost completely frozen by crosslinking, 
if its characteristic size is larger than $\xi_c$. 
The memory of the initial macroscopic texture 
makes the mechanical response non-soft.
Spatial distribution of the 
elastic free energy is 
strongly dehomogenized by applied strain,
in contrast to the case of isotropic 
crosslinking.

The influence of crosslinking
conditions has little been discussed in
previous studies of the P-M transition,
except for a few experimental
papers~\cite{Finkelmann94,Zubarev-etal}.
K\"upfer and Finkelmann~\cite{Finkelmann94}
studied both isotropic and anisotropic 
crosslinkings under external stress 
of various magnitudes.
Fig.8 in the reference
shows that polydomain networks 
crosslinked in the 
nematic phase are harder
than those prepared in the isotropic phase.
Another example of soft and non-soft 
P-M transitions
is given in Ref.~\cite{Zubarev-etal},
where it is stated that
some of their samples were prepared
above the isotropic-nematic transition temperature of the melt,
while the others are crosslinked below it.
Unfortunately, they do not explicitly state
the crosslinking condition for 
each stress-strain curve.
We wish further effort in this direction 
to be made in the future,
especially to find more evidence of 
vanishing macroscopic stress.

A remark should be made in relation to this.
We have assumed that the quenched heterogeneities 
have mesoscopic sizes
in the case of isotropic crosslinking.
However, if the network is
crosslinked in poor solvents or near the 
spinodal line, the heterogeneities can be 
macroscopic and cause strong effective disorder.
Therefore, the mechanical response
should be discussed in terms of the
size of the heterogeneity, not only
on the phase where the gel is fabricated.
Another problem in interpretation of
strain-stress data
arises from slowness of dynamical relaxation.
A recent dynamic measurement 
by Clarke and Terentjev~\cite{Clarke-Terentjev} strongly 
suggests
that the stress level will be substantially lowered
in the final equilibrium state,
which is not reachable on a practical timescale.
It might be possible that a soft equilibrium P-M transition
is masked behind a stress plateau of a sizable height,
which is reported in earlier studies~\cite{Finkelmann89,Finkelmann94}.

We have studied dynamical relaxation
after a quench from the isotropic phase.
The structure factor develops a peak at 
a finite wavenumber, which goes to zero
as the true equilibrium is approached.
Both the inverse 
peak wavenumber and the correlation length
show a power-law type growth
in an intermediate stage, 
while the elastic free energy 
is almost completely minimized in an 
early regime of the coarsening process.

Some of the experimentally observed features
of the ``four-leaf clover'' scattering pattern
have been reproduced in the present work.
Firstly, we propose that
the finiteness of the observed peak wavenumber
is explained by the slow relaxation.
The experimental peak wavenumber does not 
change during the P-M transition~\cite{Clarke2}.
Together with our simulation result,
it suggests that the coarsening is 
very slow and does not occur in the 
timescale of observation.
Further experimental study of structural relaxation 
in conjunction with stress relaxation 
would be informative to check this point.
Secondly, the peak intensity increases
and then decreases as we stretch the gel.
Qualitatively the same monotonic behavior
is reported in the experiment.
The initial increase is due to a sharpening
the peak, which is partially understood
by the fact that the director fluctuation
at $\lambda=\lambda_m$ is soft
only on a plane and a line in the $\bq$-space.

We close by listing some open questions.
(i)
We did not answer
whether the long-range order is destroyed
by an {\it arbitrarily} weak disorder
under {\it no external stress}
or, equivalently, when the
average strain $\lambda$ is not externally
constrained.
A shift of the ground state
from the monodomain ($\lambda=\lambda_m$) 
to polydomain ($\lambda=1$) states
should occur, either gradually or abruptly, 
as we increase the disorder strength from zero.
Probably this problem is not of practical 
importance because of a small but finite hysteresis 
and the slow dynamics. 
(ii) Stretching-induced-anisotropy of the depolarized scattering pattern
as we numerically find is contrary to the experimental 
observation. We may suggest an effect of spatial dimensionality.
In three dimensions there are three Frank constants,
whose relative strengths may affect the anisotropy.
Experimental investigation of 3D domain structure 
would be informative.
(iii) Much remains to be done for understanding
dynamical relaxation to the final equilibrium state.
In theoretical part, the origin of the apparent power law 
is yet unknown.
Dynamic equations for dry elastomers are to be
constructed, taking the intra-network friction account.
In numerical part, late stages of the relaxation process is left unexplored.
Stress relaxation for strong quenched disorder
and after stretching should be addressed 
to make a comparison to experiment.
As these necessitate extensive computation, 
we leave them for future work.

\acknowledgments

The author is grateful to
Professor Akira Onuki
for helpful comments 
and discussions.
He also thanks
Professor Ken Sekimoto,
Dr. Alexandra ten Bosch,
and Dr. Jun Yamamoto 
for valuable discussions.

\appendix
\section{Effective free energy in the monodomain state}

Here we sketch the derivation of \Eq{Feffmono}.
Substituting 
Eqs.(\ref{averagestrain}), (\ref{displacement}) and (\ref{Lambda})
into \Eq{Felbasic},
we have
\eq
F_{el} &=& 
\f{\mu}{2}
\int d\br \bigglrL{
C_{ij} L_{ij} 
+ 2 C_{ik} L_{jk} (\D_i u_j)
\nn\\&& \qquad
+ \Av{C}_{ij} \Av{L}_{kl} (\D_i u_k)(\D_j u_l) 
+ \kappa (\D_i u_i)^2
},
\label{Felmono2}
\qe
where $C_{ij} = (\delta_{kl} + R_{kl}) \Av{\Lambda}_{ik} \Av{\Lambda}_{jl}$
and $L_{ij}=\delta_{ij} - \alpha Q_{ij}
= (1+\alpha/3)\delta_{ij} - \alpha n_i n_j$.
In the third term of the integrand we have replaced $C_{ij}$ and $L_{ij}$
with their spatial averages as the deviations will contribute only to
higher order terms in the effective free energy.
The last term is added by hand to
temporarily relax the incompressibility condition (\ref{incompressibility}),
which is recovered by taking the limit $\kappa \to \infty$ 
afterwards.
The condition of mechanical equilibrium 
(\ref{equilibrium}) 
can be written as
\eq
\D_i(C_{ik}L_{jl}) 
+ \Av{C}_{ij} \Av{L}_{kl} \D_i \D_j u_k + \kappa \D_i \D_j u_j = 0
\label{equilibrium2}
\qe
Taking the incompressible limit $\kappa \to \infty$, we have 
\eq
\bu(\bq) = \f{1}{\tC:\bq\bq}
\bigglrL{\tL^{-1} \cdot \bg(\bq) 
- \f{\tL^{-1} \cdot \bq}{\tL^{-1}:\bq\bq} \;
\bigglrS{
\bq\cdot \tL^{-1} \cdot \bg(\bq)
}
},
\label{uR}
\qe
where $\bg$ is an auxiliary variable defined by
\eq
g_i(\br) &=& \D_j (C_{jk} L_{ik}).
\label{define_g}
\qe
Substituting \Ref{uR} into \Ref{Felmono2}, we obtain an effective free energy,
\eq
\tilde{F}_{el} &=& \f{\mu}{2}\int d\br \tC : \tL
\nn\\
&+& \f{\mu}{2}\int_{\bq} \f{1}{\Av{\tC} : \bqh\bqh}
\bigglrL{
\f{1}{\Av{\tL}^{\;-1}:\bqh\bqh} 
\biglrF{\bqh \cdot \Av{\tL}^{\;-1} \cdot \bg(\bq)}^2 
\nn\\&&\hspace{20mm}
- \bg(\bq) \cdot \Av{\tL}^{\;-1} \cdot \bg(-\bq)
},
\label{Felmono3}
\qe
We arrive at \Eq{Feffmono}
by putting 
\Ref{averagestrain} into $C_{ij}$,
\Ref{deltan}
into $L_{ij}$, and the
resulting expressions
into \Ref{define_g} 
and \Ref{Felmono3}.

\end{multicols}
\begin{multicols}{2}
\begin{figure}
\myfigure{
\epsfxsize=115pt \epsffile{fig3.eps}
}
{
Preferred local director configuration.
Ellipses indicate the anisotropy of local strain.
The ``checkered'' structure enables
each domain to elongate along the local director,
without violating the global constraint $\lambda=1$.
Note that the proper elastic interaction does not select
the characteristic size of the pattern. 
\label{preferN2D}
}
\end{figure}
\end{multicols} 
\begin{figure}
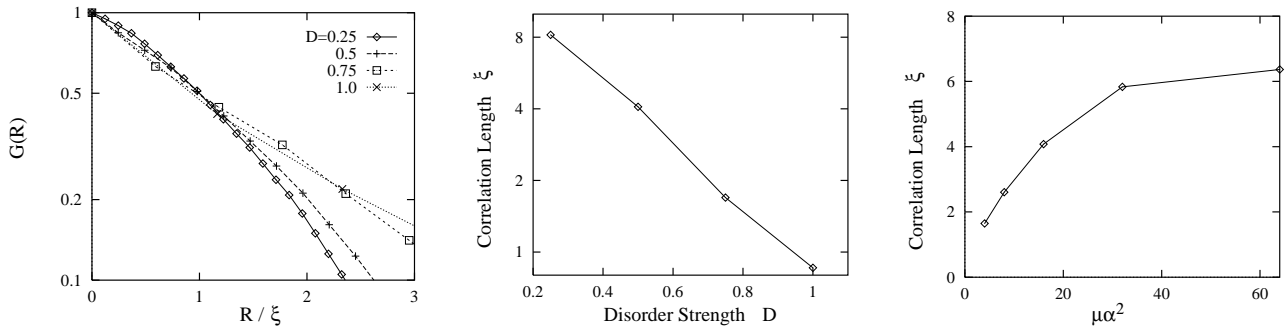

\subfigure{\epsfxsize=168pt \epsffile{fig4left.eps}}
\subfigure{\epsfxsize=160pt \epsffile{fig4middle.eps}}
\subfigure{\epsfxsize=160pt \epsffile{fig4right.eps}}
\caption{
Left:
Correlation function $G(R)$
for different disorder strengths.
\label{corrG}
Middle:
Correlation length versus disorder strength.
In the left and middle plots we fix $\mu\alpha^2=16$.
Right:
Correlation length
versus strength of elastic interaction.
The disorder strength is fixed ($D=0.5$).
\label{Xi}
}
\end{figure}
\begin{figure}
\hspace{-10mm}
\subfigure{\epsfxsize=246pt \epsffile{fig5left.eps}}
\hspace{-20mm}
\subfigure{\epsfxsize=172pt \epsffile{fig5middle.eps}}
\subfigure{\epsfxsize=130pt \epsffile{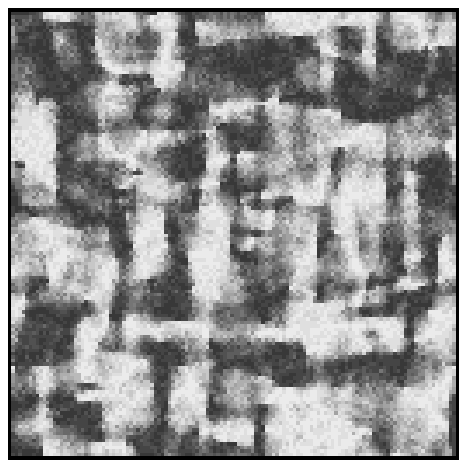}}
\caption{
Left:
Director-relative correlation function $H(R)$.
Plotted in the region $\xi<R<5\xi$.
Middle:
Depolarized light scattering intensity $\lrA{\lrF{Q_{xy}({\bf q})}^2}$.
Right:
Snapshots of the orientational order parameter field $Q_{xy}({\bf
r})=\sin2\theta$.
The value of $Q_{xy}$ is positive in
white regions and negative in black regions.
\label{corrT}
}
\end{figure}
\hspace{0mm}\vspace{-15mm}\\
\begin{multicols}{2}
\begin{figure}
\subfigure{\epsfxsize=164pt \epsffile{fig6top.eps}}
\nopagebreak[4]
\vspace{-7mm}\\
\subfigure{\epsfxsize=202pt \epsffile{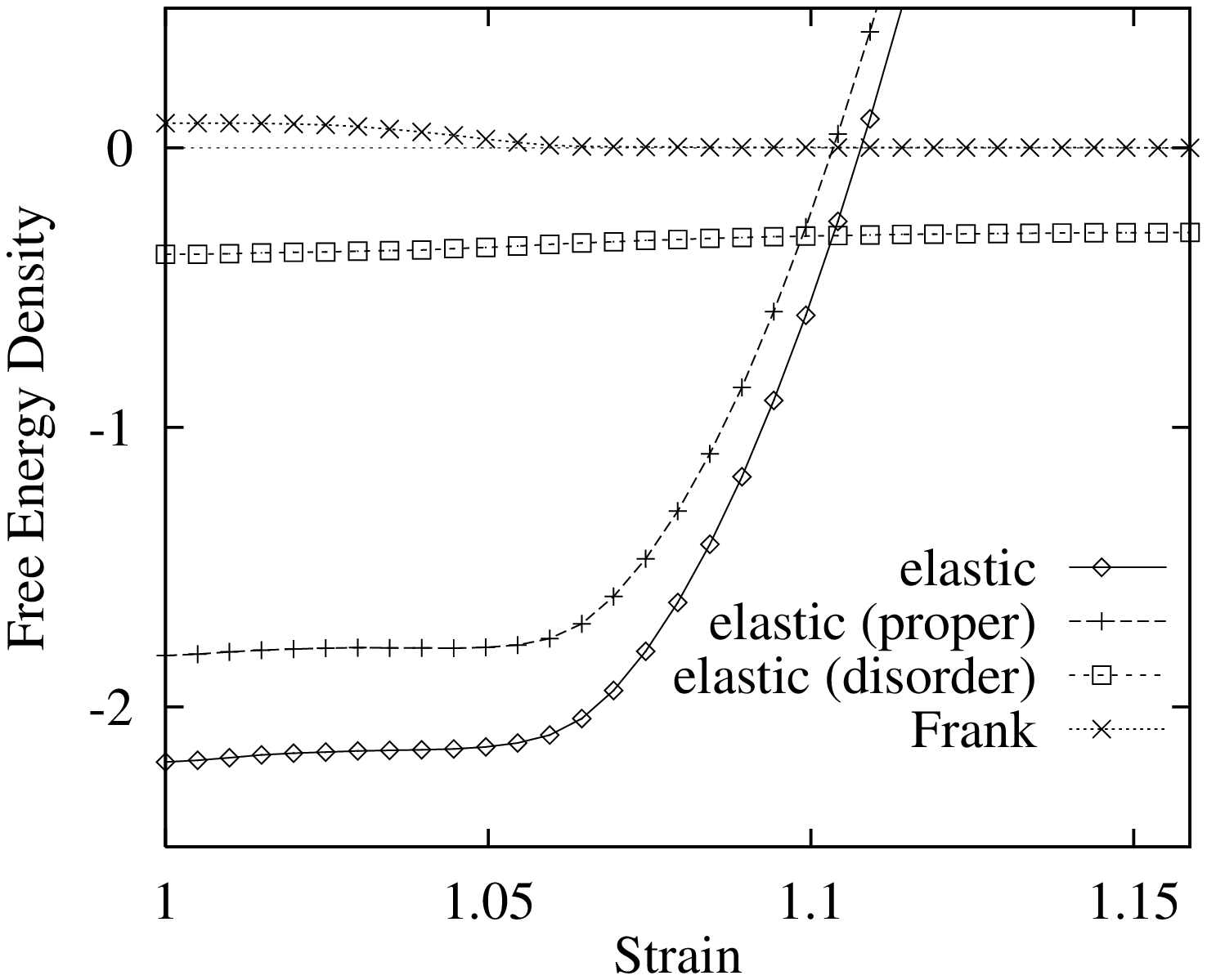}}
\mycaption{
Top:
Strain-stress ($\lambda$ vs. $\mu^{-1}\sigma_{macro}$)
and strain-orientation ($\lambda$ vs. $S$)
curves for $\alpha=0.2$ ($\lambda_m=1.051$).
Diamonds ($\gnuplotdiamond$) and crosses ($+$)
are for the loading and unloading processes, respectively.
Bottom:
Free energy densities (for the loading process).
\label{stress_orientation}
}
\end{figure}
\begin{figure}
\epsfxsize=164pt \epsffile{fig7.eps}
\mycaption{
Strain-stress and strain-orientation curves
for $\alpha=0.2,0.4,0.8$ and $1.2$ from the left to right.
Corresponding values of $\lambda_m$ are
$1.05,1.11,1.24$ and $1.41$, respectively.
The parameters $\mu \alpha^2=16$ and $D=0.5$  
are common to all cases.
\label{stress_orientation_alpha}
}
\end{figure}
\begin{figure}
\epsfxsize=202pt \epsffile{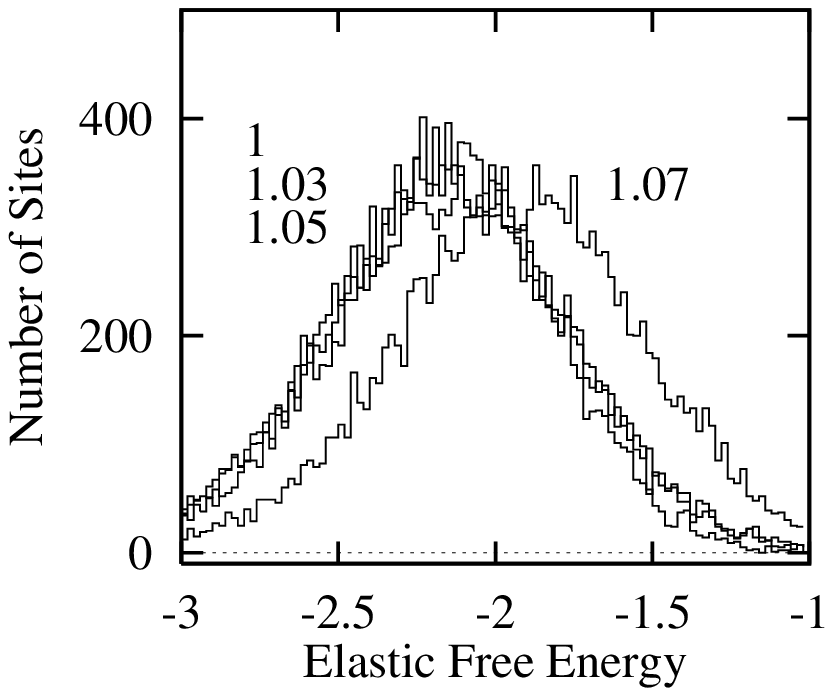}
\mycaption{
Histogram of the elastic free energy contained in a site.
Distributions for cases $\lambda=1$,$1.03$,$1.05$ and $1.07$ are shown.
The first three are indistinguishable from each other.
\label{histogram}
}
\end{figure}
\begin{figure}
\epsfxsize=204pt \epsffile{fig9.eps}
\mycaption{
Real space snapshot of the field $Q_{xy}({\bf r})$.
\label{domain_morphology}
}
\end{figure}
\end{multicols}
\vspace{-30mm}
\begin{figure}
\epsfxsize=510pt \epsffile{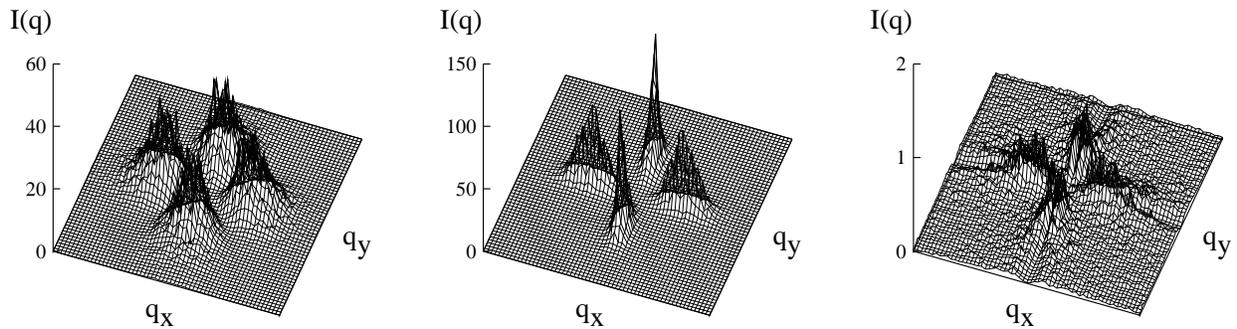}
\caption{
Depolarized scattering intensity at $\lambda=1$,$1.03$, and $1.05$
from the left to right. 
Statistical average over 20 samples is taken for each case.
Note the difference of intensity scales.
\label{scattering_intensity}
}
\end{figure}
\begin{figure}
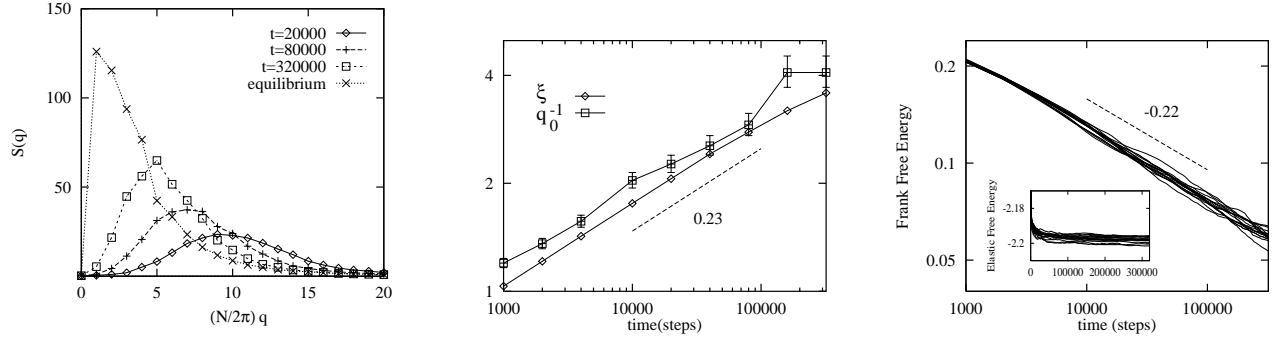

\subfigure{\epsfxsize=164pt \epsffile{fig11left.eps}}
\subfigure{\epsfxsize=164pt \epsffile{fig11middle.eps}}
\subfigure{\epsfxsize=164pt \epsffile{fig11right.eps}}
\caption{
Left:
Circularly averaged structure factor $S(q)$
at $t=2\times 10^4$,$8\times 10^4$,$32\times10^4$ steps
and in mechanical equilibrium.
Statistical average over 20 samples is taken.
Middle:
Evolution of 
the correlation length $\xi$
and inverse peak wavenumber $q_{0}^{-1}$.
\label{relaxationSq}
Right:
Dynamic relaxation of Frank and elastic free energies.
Plotted are data from 20 individual runs.
\label{relaxationEnergy}
}
\end{figure}
\begin{multicols}{2}
\begin{figure}
\epsfxsize=258pt \epsffile{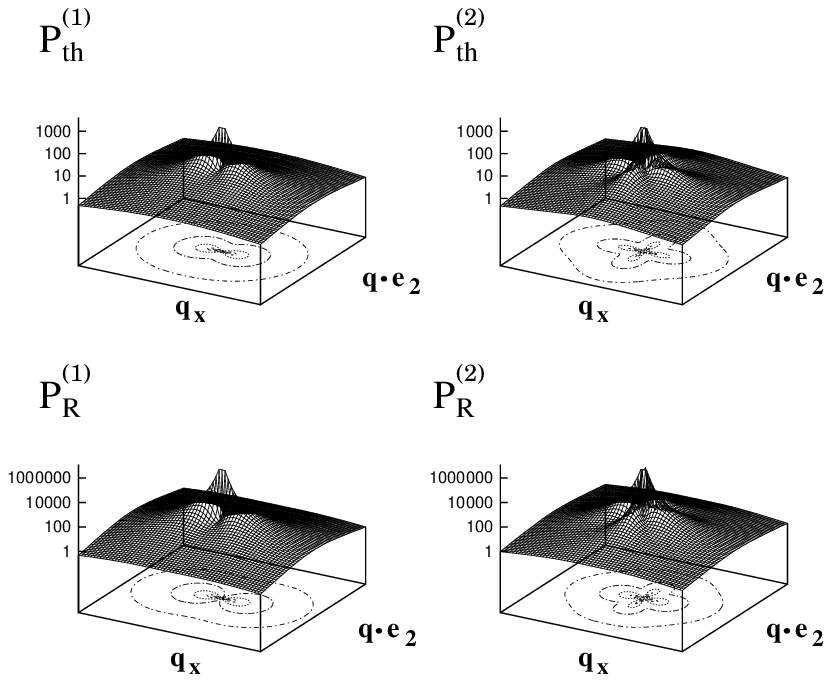}
\mycaption{
Square amplitudes of director fluctuation
(in arbitrary unit).
$P_R^{(1)}$ and $P_R^{(2)}$ are
multiplied by a common prefactor.
Shown is the region 
$-30 < K'^{1/2} q_x < 30$ and
$-30 < K'^{1/2} {\bf q} \cdot {\bf e_2} < 30$.
\label{ozdirt}
}
\end{figure}
\begin{figure}
\epsfxsize=164pt \epsffile{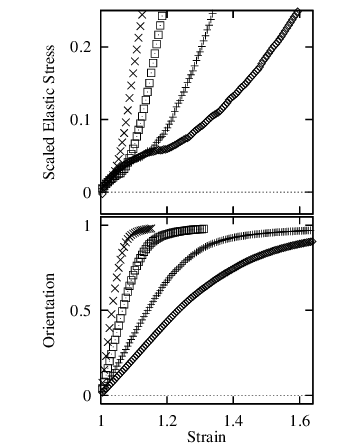}
\mycaption{
Strain-stress and strain-orientation curves
for the case of anisotropic crosslinking.
Plotted for cases 
$\alpha=0.2,0.4,0.8$, and $1.2$ 
from the left to right,
with $\mu \alpha^2=16$ in common.
\label{stress_orientation_alpha_aniso}
}
\end{figure}
\begin{figure}
\epsfxsize=204pt \epsffile{fig14.eps}
\mycaption{
Snapshots of the field $Q_{xy}({\bf r})$
for the case of anisotropic crosslinking.
The director texture of
the nematic liquid just before crosslinking
is retained at $\lambda=1$.
\label{domain_morphology_aniso}
}
\end{figure}
\begin{figure}
\myfigure{
\epsfxsize=204pt \epsffile{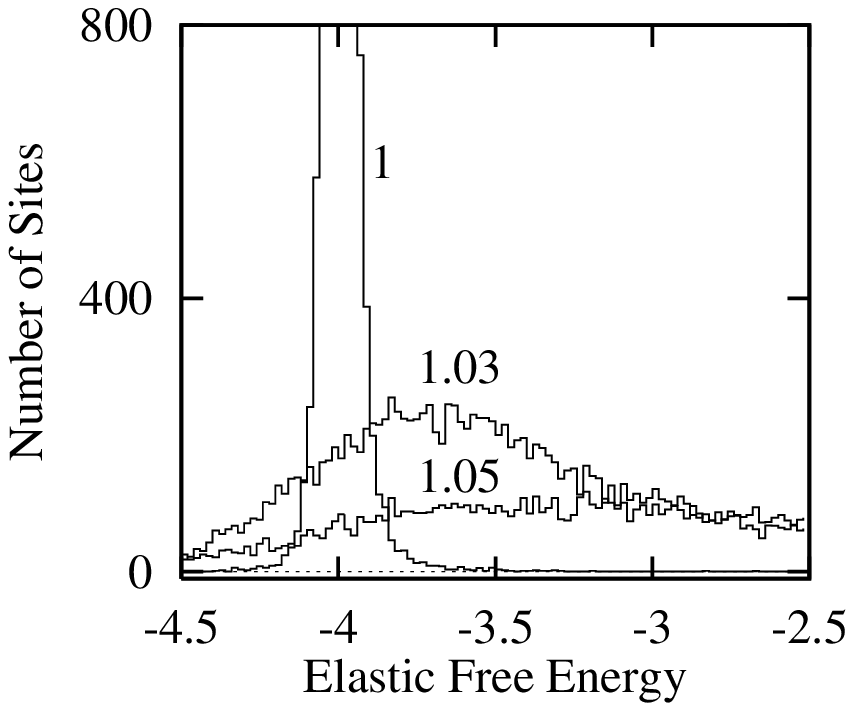}
}
{
Histogram of the elastic free energy contained in a site.
Cases $\lambda=1$,$1.03$ and $1.05$ are shown.
The peak for $\lambda=1$ 
is out of the window and
counts more than $3000$ sites.
\label{histogram_aniso}
}
\end{figure}
\end{multicols}
\end{document}